\documentclass[useAMS,usenatbib]{mn2e}
\usepackage{graphicx}
\usepackage{hyperref}
\usepackage{tabularx}
\usepackage{mathtools}
\usepackage{amsmath,amssymb}
\usepackage{chngcntr}

\newcommand{\mnras}{MNRAS}
\newcommand{\apj}{ApJ}
\newcommand{\aj}{AJ}
\newcommand{\apjs}{ApJS}
\newcommand{\nat}{Nature}
\newcommand{\apjl}{ApJL}
\newcommand{\aap}{A\&A}

\newcommand{\SBcgs}{\mathrm{erg\,s^{-1}\,cm^{-2}\,arcsec^{-2}}}
\newcommand{\lya}{Ly$\alpha$}
\newcommand{\NHI}{$\mathrm{N_{HI}}$}
\newcommand{\G}{$\mathrm{\Gamma_{HI}}$}
\newcommand{\cf}{covering fraction}

\newcommand{\flls}{$\mathrm{f_{LLS}}$}

\newcommand{\refsim}{\textit{Recal-L0025N0752} }

\title[UV Background at $z>3$]{Constraining the cosmic UV background at $z>3$ with MUSE Lyman-$\alpha$ emission observations}\author[Sof\'ia G. Gallego et al.]{ Sofia G. Gallego\thanks{E-mail:
\href{mailto:gallegos@phys.ethz.ch}{gallegos@phys.ethz.ch}, \href{mailto:sofiag.gallego@gmail.com}{sofiag.gallego@gmail.com}}$^{1}$, Sebastiano Cantalupo$^{1,2}$, Saeed Sarpas$^{1}$, Bastien Duboeuf$^{1,3}$,\newauthor\ Simon Lilly$^{1}$, Gabriele Pezzulli$^{1,4}$, Raffaella Anna Marino$^{1}$, Jorryt Matthee$^{1}$,
\newauthor\ Lutz Wisotzki$^{5}$, Joop Schaye$^{6}$, Johan Richard$^{7}$, Haruka Kusakabe$^{8}$ \&\newauthor\ Valentin Mauerhofer$^{8,7}$\\
$^{1}$Department of Physics, ETH Z\"urich, CH-8093 Z\"urich, Switzerland\\
$^{2}$Department of Physics, University of Milan Bicocca, Piazza della Scienza 3, 20126 Milano, Italy\\
$^{3}$\`Ecole Normale Sup\'erieure de Paris-Saclay\\
$^{4}$Kapteyn Astronomical Institute, University of Groningen, Landleven 12, 9747 AD Groningen, The Netherlands\\
$^{5}$Leibniz-Institut f\"ur Astrophysik Potsdam (AIP), An der Sternwarte 16, D-14482 Potsdam, Germany\\
$^{6}$Leiden Observatory, Leiden University, PO Box 9513, NL-2300 RA Leiden, the Netherlands\\
$^{7}$Univ Lyon, Univ Lyon1, Ens de Lyon, CNRS, Centre de Recherche Astrophysique de Lyon UMR5574, F-69230 Saint-Genis-Laval, France\\
$^{8}$Observatoire de Gen\`eve, Universit\'e de Gen\`eve, 51 Chemin de P\'egase, 1290 Versoix, Switzerland}
\begin{document}

\date{\today}

\pagerange{\pageref{firstpage}--\pageref{lastpage}} \pubyear{2016}

\maketitle

\label{firstpage}

\begin{abstract} 
The intensity of the Cosmic UV background (UVB), coming from all sources of ionizing photons such as star-forming galaxies and quasars, determines the thermal evolution and ionization state of the intergalactic medium (IGM) and is, therefore, a critical ingredient for models of cosmic structure formation. Most of the previous estimates are based on the comparison between observed and simulated Lyman-$\alpha$ forest.
We present the results of an independent method to constrain the product of the UVB photoionisation rate and the covering fraction of Lyman limit systems (LLSs) by searching for the fluorescent Lyman-$\alpha$ emission produced by self-shielded clouds. Because the expected surface brightness is well below current sensitivity limits for direct imaging, we developed a new method based on three-dimensional stacking of the IGM around Lyman-$\alpha$ emitting galaxies (LAEs) between $2.9<z<6.6$ using deep MUSE observations. Combining our results with covering fractions of LLSs obtained from mock cubes extracted from the EAGLE simulation, we obtain new and independent constraints on the UVB at $z>3$ that are consistent with previous measurements, with a preference for relatively low UVB intensities at $z=3$, and which suggest a non-monotonic decrease of $\mathrm{\Gamma_{HI}}$ with increasing redshift between $3<z<5$. This could suggest a possible tension between some UVB models and current observations which however require deeper and wider observations in Lyman-$\alpha$ emission and absorption to be confirmed. Assuming instead a value of UVB from current models, our results constrain the covering fraction of LLSs at $3<z<4.5$ to be less than 25\% within 150 kpc from LAEs.

\end{abstract}

\begin{keywords}
(galaxies:) intergalactic medium, (cosmology:) diffuse radiation, (cosmology:) large-scale structure of Universe, galaxies: haloes, ultraviolet: general
\end{keywords}

\section{Introduction}

The evolution of the ionization state of baryonic matter in the Universe is largely determined by the Cosmic UV Background (UVB), the average bulk of ionizing photons coming from star-forming galaxies and active galactic nuclei (AGN). In particular, these sources are  responsible for the reionization of hydrogen and helium in the Intergalactic Medium (IGM) and for sustaining its highly ionized state up to the current epoch
\citep[see][and references therein]{Meiksin2005}.
The epoch of reionization can be understood by studying the redshift evolution of the UVB right after this period. Current predictions require galaxies at the reionization era to be much more efficient at emitting ionizing photons than their low redshift counterparts. The efficiency of these galaxies is determined by their ionizing escape fraction ($f_{\mathrm{esc}}$) and their UV spectral shape. The UVB also regulates the temperature of the IGM \citep[][]{Miralda1994} and thence imposes a temperature threshold for galaxy formation in low-mass halos \citep[][]{Gnedin2000}. Moreover, given that at least half of the baryonic material is predicted to be in the IGM \citep[][]{Peebles1975,Bond1996} and that galaxies cannot sustain their star formation rate (SFR) through cosmic time with their current gas content, a continuous accretion from the IGM is a necessary ingredient for galaxy evolution \citep[e.g.][]{Keres2005, Fumagalli2011}. The UVB affects this accretion by changing the thermal and ionization state of the gas, and therefore it plays a major role in cosmic structure and galaxy formation processes \citep[e.g.][]{Efstathiou1992,Wiersma2009,Cantalupo2010}. 

Determining the intensity of the UVB at high redshift by directly measuring the emission from individual sources of ionizing photons is, however, an extremely challenging task. Many of the ionizing sources are too faint to be detected individually and part of the ionizing spectrum is highly absorbed by the IGM itself and by dust. Most of the previous studies have been focused instead on estimating the UVB indirectly by studying the properties of the IGM that are affected by the UVB photoionization rate of neutral hydrogen ($\mathrm{\Gamma_{HI}}$). 
The first constraints on the high redshift IGM came from the study of the $\mathrm{Lyman-\alpha}$ ($\mathrm{Ly\alpha}$) absorption features, in particular the so called ``\lya\ forest" in the spectra of distant quasars \citep{Lynds1971}. Early methods investigating the value of the UVB relied on measuring the redshift evolution of the \lya\ transmission in the quasar surroundings \citep[e.g.][]{Bajtlik1988,Kulkarni1993,Srianand1996,Scott2000,DallAglio2008b,Calverley2011,Wyithe2011,Romano2019}, which depends on the escape fraction and mean free path of ionizing photons from the quasar and the quasar spectra. Such methods are affected by possible anisotropies of the quasar emission and by the fact that quasar environments are likely denser compared to those of typical galaxies. Other methods compare the \lya\ forest data with cosmological simulations \citep[e.g.][]{Gaikwad2016} which depend on assumptions on the gas temperatures and the opacity of the IGM to \lya\ and ionizing photons. Theoretical models of the UVB \citep{Haardt2001,Faucher-Giguere2009,Haardt2012,Khaire2019} are constructed by solving the radiative transfer equation through cosmic times, and depend on a set of input parameters such as the evolution of the escape fraction of ionizing radiation ($f_{\mathrm{esc}}$), 
the spectral shape of the UVB, galaxy and AGN luminosity functions, the distribution in redshift and column density of the IGM absorbers, among other ingredients \citep[e.g.][]{Wyithe2011,Calverley2011,Becker2013,DAloisio2018}.

Another way to estimate the intensity of the UVB seen by the IGM is through \lya\ emission.  When a hydrogen atom is ionized, it can recapture a free electron (recombination) and transition into the ground state by a succession of photon emissions, most of which ends up with \lya. 
A good approximation is that the majority of \lya\ photons are produced in optically thick clouds, defined as Lyman Limit Systems (LLS) with column densities ($\mathrm{N_{HI}}$) above $10^{17.2}\mathrm{cm}^{-2}$, corresponding to an optical depth at the Lyman Limit ($\mathrm{\tau_{LL}}$) above 1 \citep{Gould1996}. 
This process is called fluorescent \lya\ emission, and in the absence of other \lya\ emission mechanisms the expected surface brightness (SB) is directly proportional to the intensity of $\mathrm{\Gamma_{HI}}$ \citep{Miralda1990,Gould1996,Bunker1998,Cantalupo2005}.

 For $z>3$, $\mathrm{SB_{Ly\alpha}}$ is predicted to be about $10^{-20}$ $\SBcgs\ $ \citep{Gould1996,Furlanetto2005,Cantalupo2005,Bertone2012,Witstok2019}, well below current observational limits on individual detections. One way to enhance the \lya\ emission is to look in the vicinity of quasars and AGN which increase the intensity of the ionizing radiation up to very large radii \citep{Cantalupo2014,Battaia2019,Umehata2019}. The detected emission is in this case independent on the value of the UVB and therefore these methods cannot be used to provide a constraint on the UVB.  
 Therefore, the only viable method with current observations to increase the detectability of the UVB fluorescent emission requires the stacking of deep observations of average regions of the universe where the fluorescent signal is expected to be present.  
 In our previous work \citep[][]{Gallego2018} we developed a novel ``oriented-stacking" technique to detect the \lya\ fluorescent signal of the IGM around \lya\ emitting galaxies (LAEs), tailored to highlighting the filamentary structure of the Cosmic Web.  By stacking deep MUSE observations \citep{Bacon2015,Bacon2017}, using \lya\ emitting galaxies (LAEs) as reference points for the three-dimensional orientation of each stacking element, extracting subcubes around them and transforming the coordinates of the field to align it with a neighbouring LAE, we reached a $2\sigma$ SB limit of $0.4\times10^{-20}\SBcgs,\ $ three times below expectations from \citet[][hereafter HM12]{Haardt2012}, under the assumption that all of the selected pairs of LAEs were connected by LLS filaments.
 This, however, does not necessarily imply that the UVB is inconsistent with HM12; in fact, the largest source of uncertainty of this technique is the unknown fraction of LLS filaments between galaxy pairs, which will depend on the galaxies' properties but especially on the distance between the selected pairs. Restricting the sample of galaxy pairs to the closest ones (e.g. below 2 cMpc) is not sufficient to compensate for the increased noise associated to a reduction of the sample size. 
 
 In this paper, we instead perform 3-D radial stacking of LAEs to estimate the value of the UVB and specifically $\mathrm{\Gamma_{HI}}$. Although we lose any information on the 2-D spatial distribution of LLSs and the possible detection of cosmic filaments, the decrease in the average noise that results from stacking a larger number of pixels is sufficient to compensate for the decrease in the expected fluorescent signal.
Because the expected signal in our stacking analysis depends both on the UVB intensity and the LLS covering fraction, 
our method can provide a constraint on the UVB only if the covering fraction of LLS around galaxies is known. 
Despite this limitation, we notice that this methodology does not require several of the assumptions made by
previous studies and it could therefore provide an independent constraint on the value of the UVB.
Given the lack of observational constraints on \flls\ around LAEs at the redshifts available for MUSE, we decided to estimate \flls\ from mock cubes obtained from the EAGLE simulation \citep{Schaye2015}. However, we emphasize that our observational results are independent of the estimated \flls\ from simulations, and therefore they can be used to estimate \G\ based on future studies of $\mathrm{f_{LLS}}$. Moreover, our observational results may also be used as a prediction for the radial distribution of LLSs around galaxies given an assumed value of \G.

This paper is organized as follows: Section \ref{sec:uvb} describes the theoretical background on the estimation of \G\ from the expected IGM \lya\ emission, Section  \ref{sec:observations} the observational data, data reduction, the galaxy catalog and stacking procedure, and Section \ref{sec:simulations} the mock cubes produced to estimate $\mathrm{f_{LLS}}$ and the uncertainties in the estimation. 
Results on the value of \G\ and the \cf\ of LLS from combining observations and simulations are shown in Section \ref{sec:results}. A discussion on the assumptions on the calculation of \G\ and \flls\ and the implications for future observations are presented in Section \ref{sec:discussion} and a summary is presented in Section \ref{sec:summary}. Throughout the paper we assume a flat $\Lambda$CDM cosmology with $\mathrm{H_0}=69.6\,\mathrm{km\,s^{-1}Mpc^{-1}}$, $\mathrm{\Omega_m}=0.286$, $\mathrm{\Omega_\Lambda}=0.714$ \citep{Bennett2014} and $\mathrm{\Omega_b}\,h^{2}=0.02223$ \citep{Bennett2013}.

\section{Fluorescent \lya\ emission from the UVB}\label{sec:uvb}
In this section we develop the formalism needed to calculate the expected \lya\ SB produced by IGM clouds photoionized by the UVB, first developed in \citet{Gould1996}.

The rate at which the UVB photoionizes the neutral hydrogen (\G) is denoted as:

\begin{equation}
    \label{gamma}
    \mathrm{\Gamma_{HI}} = 4\pi\int^{\infty}_{\nu_0} \dfrac{J_{\nu}\,\sigma_\nu\,d\nu}{h\nu},
\end{equation}
where $J_{\nu}$ is the angled-averaged monochromatic intensity (per unit time, frequency, area and solid angle), and $\sigma_\nu$ is the hydrogen photo-ionization cross section. The integral lower limit is defined by the hydrogen ionization threshold ($\nu_0=\nu_{\mathrm{HI}}$, $h\nu_{\mathrm{HI}}=13.6$ eV) and the upper limit is in principle infinity but in practice it is truncated at $\nu_{\mathrm{HeII}}=4\nu_0$ given that most of the radiation above that threshold is absorbed by helium and in general it contributes very little to the overall absorbed energy \citep{Cantalupo2005}. 

In a self-shielded cloud the incident radiation will ionize the outermost layer and a fraction of the recombination radiation will be emitted as \lya\ photons. The \lya\ photon production per unit time, surface and solid angle can be expressed as:

\begin{equation}\label{lyarate}
  \mathrm{R_{Ly\alpha}}   =  \mathrm{\epsilon_{thick}} \int^{4\nu_0}_{\nu_0} \dfrac{J_{\nu}\,d\nu}{h\nu},
\end{equation}
where $\mathrm{\epsilon_{thick}}$ is the fraction of recombinations leading to a \lya\ photon, a factor that depends weakly on the temperature and varies between 0.61 to 0.68 for temperatures between $10^4$K to $10^5$K and case B recombination.

If we approximate \G\ as:
\begin{equation}
    \label{gamma2}
    \mathrm{\Gamma_{HI}} \simeq 4\pi\,\bar{\sigma}_{\nu_\mathrm{HI}} \int^{4\nu_0}_{\nu_0} \dfrac{J_{\nu}\, d\nu}{h\nu},
\end{equation}
where $\bar{\sigma}_{\nu_\mathrm{HI}} \equiv \dfrac{\int^{4\nu_0}_{\nu_0}J_{\nu}/\nu\,\sigma_\nu d\nu}{\int^{4\nu_0}_{\nu_0}J_{\nu}/\nu\, d\nu}$ is the grey cross section,
we can rewrite Eq. \ref{lyarate} as

\begin{equation}
  \mathrm{R_{Ly\alpha}}   \simeq   \dfrac{\mathrm{\epsilon_{thick}}\,\mathrm{\Gamma_{HI}}}{4\pi\,\bar{\sigma}_{\nu_\mathrm{HI}}}.
\end{equation}
$\mathrm{R_{Ly\alpha}}$ can be transformed into the observed surface brightness (SB):

\begin{equation}
\mathrm{SB_{Ly\alpha}} \simeq  \dfrac{\mathrm{R_{Ly\alpha}}\times h\nu_{Ly\alpha}}{(1+z)^4},
\end{equation}
where the factor $(1+z)^4$ accounts for the cosmological SB dimming.

Assuming $\mathrm{\epsilon_{thick}}=0.65$ for $\mathrm{T=2\times10^4\,K}$, the expected equilibrium temperature for the photoionized self-shielded gas layers \citep{Cantalupo2005}, we estimate a \lya\ SB of:


\begin{equation}
\begin{aligned}
\mathrm{SB_{Ly\alpha}} \simeq 1.84\times10^{-17} \SBcgs \\ 
\left(\dfrac{\mathrm{\Gamma_{HI}}}{10^{-12}\mathrm{s^{-1}}}\right)
\left(\dfrac{\bar{\sigma}_{\nu_\mathrm{HI}}}{10^{-18}\mathrm{cm^2}}\right)^{-1}
(1+z)^{-4}.
\end{aligned}
\end{equation}

Using the HI ionization cross section from Osterbrock (1974), the grey cross section $\bar{\sigma}_{\nu_\mathrm{HI}}$ will depend only on the spectral shape of the UVB and therefore we expect $\mathrm{SB_{Ly\alpha}}$ to be proportional to the photoionization rate. Assuming that the shape of the ionizing spectra does not change significantly between UVB models, the observed value of \G\ will be a simple rescaling of the observed SB (or upper limit) with respect to the SB expected for a given model:

\begin{equation}\label{maxSB}
\begin{aligned}
\mathrm{\Gamma_{HI}} \simeq 0.54\times10^{-12}\mathrm{s^{-1}} \left(\dfrac{\mathrm{SB_{Ly\alpha}}}{10^{-17}\SBcgs}\right) \\ \left(\dfrac{\bar{\sigma}_{\nu_\mathrm{HI}}}{10^{-18}\mathrm{cm^2}}\right)
(1+z)^{4}.
\end{aligned}
\end{equation}

\begin{equation}\label{rescale}
   \mathrm{\Gamma_{HI} = SB_{Ly\alpha, obs}\times\dfrac{\Gamma_{HI, model}}{SB_{Ly\alpha, model}}}.
\end{equation}

For the rest of the paper, we choose HM12 as our reference model, which for a value of the UVB of \G\ $=0.7\times10^{-12}\mathrm{s^{-1}}$ at $z=3.5$ we predict a SB of $1.14\times10^{-20}\SBcgs\ $\citep[][]{Gallego2018}.

\subsection{The covering fraction of Lyman limit systems}\label{sec:flls}

As discussed previously, equation \ref{rescale} is valid for the Lya SB arising from a single LLS. However we expect in general that only a fraction of a considered observed area would be covered by LLSs. Furthermore, some of the  emission may arise from optically thin clouds as well. For simplicity, we encapsulate both effects in one parameter \flls\ (for a discussion of this approximation see Section \ref{self}), so that:

\begin{equation}\label{gcalc}
   \mathrm{f_{LLS} \times \Gamma_{HI}= SB_{Ly\alpha, obs}\times\dfrac{\Gamma_{HI, model}}{SB_{Ly\alpha, model}}}.
\end{equation}

From equation \ref{gcalc}, we can infer either the average \cf\ of LLSs within the area where the SB is measured, by assuming a measured value of \G, or we can predict \G\ by assuming a value of \flls.


\section{Observations}\label{sec:observations}

The Multi-Unit Spectroscopic Explorer (MUSE, Bacon et al. 2010), mounted on the Very Large Telescope (VLT) at the Paranal Observatory, is uniquely suited for the search of extended \lya\ emission at high redshift. MUSE is a panoramic integral-field spectrograph with a field-of-view (FOV) of $1\times1$ arcmin$^2$ and a sensitivity in the optical range ($\mathrm{470 nm<\lambda<940 nm}$), capable to detect Lyman alpha emission for $2.9<z<6.6$.  The MUSE voxel (3-D pixel) size of $0.2"\times0.2" \times 1.25\,\mathrm{\text{\AA}}$ is equivalent to a cube with a physical size of about $\mathrm{1.5 kpc\times 1.5\,kpc\times\,120 \,kpc}$ at redshift 3.5.

 In this paper we use the MUSE Ultra Deep Field (UDF, Bacon et al. 2017), obtained during the Guaranteed Time Observations of the MUSE Consortium. The UDF consists of a mosaic of nine 10 h exposure fields with a total FOV of 3$\times$3 arcmin$^2$ (hereafter UDF-mosaic), plus one overlapping 31 h exposure in a 1.15 arcmin$^2$ field (hereafter UDF-10). The UDF-mosaic and UDF-10 achieve a $2\sigma$ emission line SB limit at $5500\mathrm{\text{\AA}}$ for an aperture of 1 arcsec$^2$ of about 1.3 and 0.8 $\times 10^{-19} \mathrm{erg\,s^{-1}\,cm^{-2}, arcsec^{-2}}$, respectively.

\subsection{Data reduction}

UDF-10 and UDF-mosaic data-cubes have been reduced following the procedure described in Bacon et al. (2017) and Conseil et al. (2016). 
For the purposes of detection of \lya\ emission up to large distances from galaxies and to reduce any low level systematics enhanced during stacking, we have extended the process of background correction.

The standard sky subtraction process assumes that astronomical objects have well defined boundaries with respect to the background noise. In the case of faint halos around galaxies, that extend below the noise level, this implies that some of the emission will be considered sky and therefore subtracted, especially in wavelengths with a high density of galaxies. This effect, although in general very small, is enhanced when performing stacking procedures and, for our case, its magnitude is about the same as the expected signal.

In order to correct for the over-subtraction, we recalculate the background layer by layer in the wavelength direction. Firstly, we mask continuum sources with CubEx (Cantalupo in prep.) on the white-light image.
Secondly, we mask a region of 20'' radius and spectral width of $12.5\mathrm{\text{\AA}}$ around the 3-D peak of the detected LAEs, where we expect most of the extended \lya\ emission to be present. We estimated the 3$\sigma$-clipped average flux of the unmasked voxels and subtracted it from the layer flux. 
On the background corrected data-cubes, we apply a continuum subtraction using a median filter with a wavelength width of 20 layers and a smoothing radius of 2 layers. Wavelengths with skylines are discarded.

\subsection{Galaxy Catalog}
 
\begin{figure}
\begin{center}
\includegraphics[trim={.4cm 0cm .1cm 0cm},clip,width=3.3in]{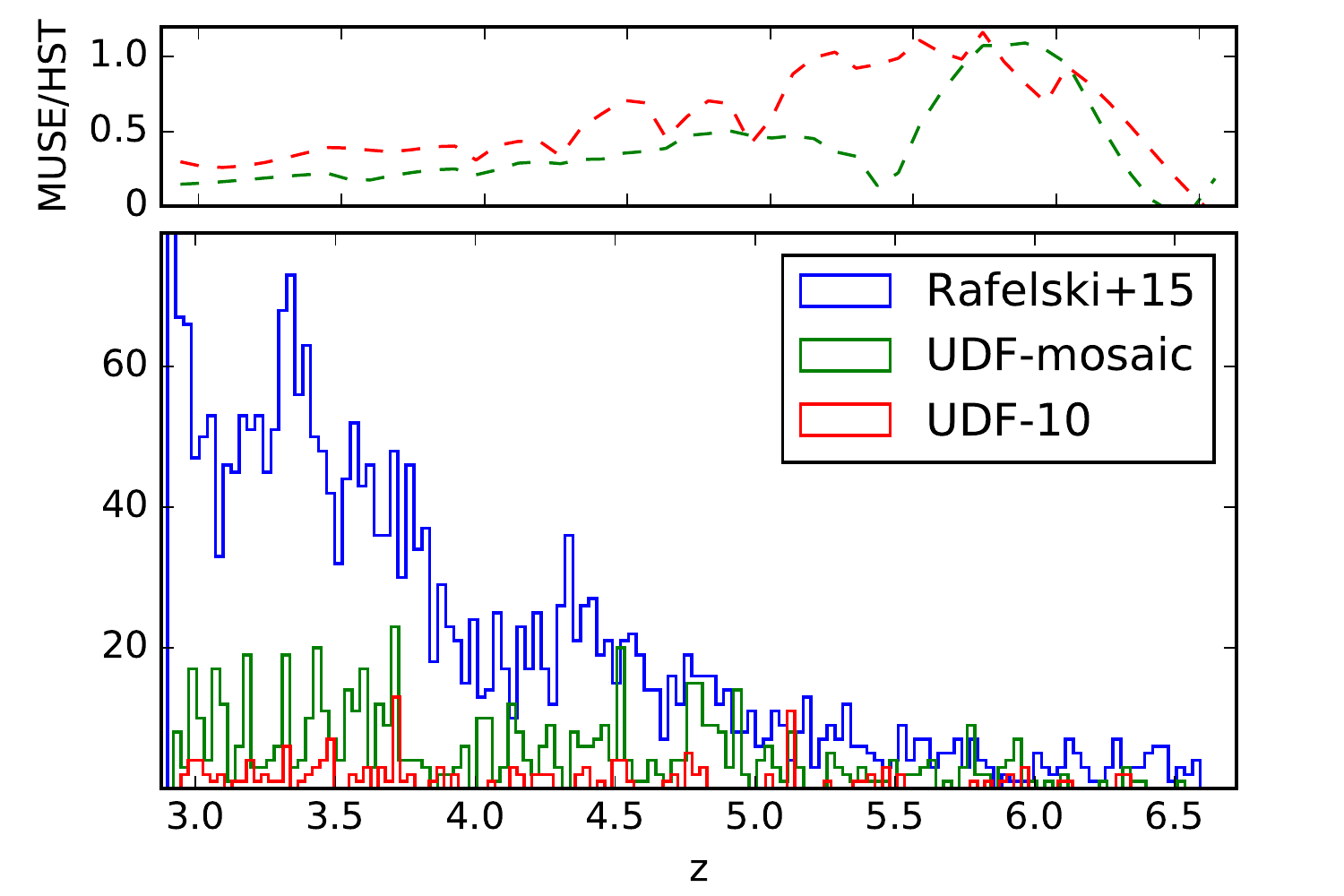}
   
   \caption{Lower panel: LAE redshift distribution for the UDF-mosaic and UDF-10 MUSE spectroscopic catalogs and the \citet{Rafelski2015} HST photometric galaxy catalog. 
Top panel:  Ratio of the UDF-mosaic and UDF-10 LAEs with respect to the HST catalog (some regions are above 1 since more sources were detected with MUSE). }\label{laehist}
\end{center}
\end{figure} 

We select high-redshift LAEs from an upgraded version of the galaxy catalogue presented in \citet{Inami2017}.
In Figure \ref{laehist} we plot the redshift distribution of LAEs in the MUSE UDF-10 and UDF-mosaic fields from our galaxy catalog, compared with the HST photometric redshift catalog from \citet{Rafelski2015} (within the UDF-mosaic field of view). The ratio between the LAEs detected in the MUSE catalogs with respect to the Hubble Space Telescope (HST) photometric catalog (in the respective fields) is calculated for 50 redshift bins (Figure \ref{laehist} top panel). Because of the advantages of MUSE IFU with respect to HST (despite the atmosphere contamination), some regions have a ratio above 1. At least one clear overdensity in the redshift distribution of LAEs is observed at about $z=3.7$, present in both the MUSE and the HST catalogs. This overdensity does not seem to artificially appear from a relative decrement in galaxy counts in their surroundings due to skylines. 
Since we expect that the local ionizing background around overdensities may be enhanced, e.g. by the presence of 
(undetected) AGN, and at the same time the \cf\ of LLS could be larger in these special volumes of the universe, 
we discarded in our main analysis LAEs within $\Delta z=0.05$ from this known overdensity in order to not bias our measurements or possible 
constraints on the UVB (see Section \ref{AGNeffect} for an analysis on this region). 

We remove LAEs from the UDF-mosaic catalog that are already present in the UDF-10 catalog and those with poor signal-to-noise (SNR, confidence level below 2 on the catalogs) or closer than 10 pixels to the border of the UDF-mosaic cube to avoid noisy pixels. With this selection criteria we end up with 138 LAEs for UDF-10 and 598 LAEs for UDF-mosaic.  

In order to maximize the SNR without extending to a redshift range where \G\ and \flls\ may significantly evolve, we divide the galaxy catalog into 4 redshift bins: $[2.9, 3.4]$, $[3.4, 4.5]$, $[4.5, 5.5]$ and $[5.5, 6.5]$, with mean redshifts 3.1, 3.9, 4.9 and 5.9, respectively.

\subsection{Subcube Extraction and Stacking}

We extract subcubes of 40"$\times$40"$\times125\,\mathrm{\text{\AA}}$ centered on the 3D peak of the \lya\ emission of the selected galaxies.
To correct for any local variations on the background noise of each subcube, we subtract the 3-$\sigma$ clipped average level of voxels at a distance larger than 20" and $12.5\,\mathrm{\text{\AA}}$ from the peak of the \lya\ emission (the center of the subcubes). This procedure is similar to the procedure adopted by Wistozki et al. (2018), where the data beyond 6" of the detected LAEs was truncated to zero, but we extend the truncation radius to much larger values since we are interested in the emission
well beyond the CGM scales. 
Since UDF-10 and UDF-mosaic cubes have different integration times, we stack the extracted subcubes for each datacube separately. We apply an average 3-$\sigma$ clipping algorithm with a single iteration for each voxel and later on combine the stacked cubes of the individual fields with a weight given by their relative depth.

\section{Simulations}\label{sec:simulations}
Given the lack of observational constraints on the distribution of HI column densities around galaxies at the redshift range of our study, in particular on the \cf\ of LLSs around LAEs,
we construct \NHI\ mock cubes based on the EAGLE project \citep{Schaye2015,Crain2015}. EAGLE is a set of cosmological simulations performed with the smoothed particle hydrodynamics (SPH) code GADGET-3 \citep{Springel2005} with a modified implementation of SPH, time stepping and subgrid models. AGN and stellar feedback are implemented with the prescription of Dalla Vecchia \& Schaye (2012).

We use for our analysis the  \textit{Recal-L0025N752} simulation, with a periodic box of 25 comoving Mpc (cMpc), a baryonic mass resolution of $2.26 \times 10^5\,\mathrm{M_\odot}$, 
and a stellar and AGN feedback implementation recalibrated to fit the galaxy stellar mass function at redshift 0.

We choose the simulation with the highest mass resolution available in order to have better resolved properties of the simulated galaxies. Massive dark matter halos are less abundant than in the larger box size simulation, although, since our stacking analysis is based on LAEs, which are believed to have halo mass below $10^{12}\,\mathrm{M_\odot}$, the lack of massive halos is not critical.
Whereas the projected area of the simulation is several times larger with a shorter line of sight distance in comparison with the MUSE fields, the total simulation volume is large enough for a comparable sample of galaxies and it allows us to extend the analysis up to a larger projected distance than the MUSE fields.

\subsection{Mock Cubes}

In order to create mock cubes from the simulation, we firstly convert the particle data of each EAGLE snapshot into a block-structured adaptively refined grid of rectangular cells using the SPH kernel to distribute the mass of each particle among its neighboring cells.

The neutral hydrogen density ($\mathrm{n_{HI}}$) for each cell is estimated by calculating the collisional and photoionization equilibrium using the on-the-spot approximation (Backer \& Menzel 1962), assuming a homogeneous photoionization rate from the UVB with values taken by the model of Haard \& Madau (\citeyear{Haardt2012}, a newer model with respect to the Haardt \& Madau \citeyear{Haardt2001} used during the EAGLE simulations).
Following the previous work of \citet{Rahmati2013}, we use a fitting function to adjust the equilibrium values to account for self-shielding and case A recombination. The hydrogen density and temperature of each cell are used for this calculation, although, since EAGLE imposes an equation of state for star-forming gas particles, we set the temperature for all star-forming particles to $10^4\,\mathrm{K}$.

Following the $\mathrm{n_{HI}}$ estimation, the adaptively refined rectangular cells are converted into a regular grid with an average of 12 particles per cell, which results in a cube of 4096$^3$ voxels. 
Finally, we convert $\mathrm{n_{HI}}$ into column densities ($\mathrm{N_{HI}}$) by collapsing one coordinate of the box into regular layers of about $1.25\,\mathrm{\text{\AA}}$. This specific width was chosen for a number of reasons: for an accurate comparison to the MUSE data that have the same wavelength width; to avoid projection effects in our $\mathrm{N_{HI}}$ calculation and to have resolution elements that are much bigger than the typical size of LLS clouds. 
A summary of all the mock cubes is presented in Table \ref{mocks}.

\begin{table}
\centering
\caption{Summary of mock cubes used in this work. The number of layers in the l.o.s. direction is selected to correspond to the MUSE wavelength width of $1.25\,\mathrm{\text{\AA}}$.}
\label{mocks}
\begin{tabular}{| >{\centering}m{1.2cm} | >{\centering}m{1.cm} | >{\centering}m{1.cm} | >{\centering}m{1.cm} | >{\centering}>{\centering}m{1.cm}@{}  m{0.2cm}@{} |}
\hline
\vspace{2mm}Snapshot & \vspace{2mm}$z$ &  \vspace{2mm}l.o.s. width [cMpc]& \vspace{2mm}$\Delta v$ [km/s] & \vspace{2mm}$\Delta z$ & \\ [4ex]
\hline    
\hline    
  \vspace{2mm}6  & \vspace{2mm}5.97 & \vspace{2mm}0.44 & \vspace{2mm}44.3 & \vspace{2mm}0.058 &\\ [2.ex]
\hline
  \vspace{2mm}7  & \vspace{2mm}5.49 & \vspace{2mm}0.50 & \vspace{2mm}47.9 & \vspace{2mm}0.052 &\\ [2.ex]
\hline
 \vspace{2mm}8  & \vspace{2mm}5.04 & \vspace{2mm}0.56 & \vspace{2mm}51.0 & \vspace{2mm}0.046 &\\ [2.ex]
\hline
 \vspace{2mm}9  & \vspace{2mm}4.49 & \vspace{2mm}0.64 & \vspace{2mm}56.9 & \vspace{2mm}0.041 &\\ [2.ex]
\hline
 \vspace{2mm}10  & \vspace{2mm}3.98 & \vspace{2mm}0.74 & \vspace{2mm}61.9 & \vspace{2mm}0.035 &\\ [2.ex]
\hline
 \vspace{2mm}11  & \vspace{2mm}3.53 & \vspace{2mm}0.86 & \vspace{2mm}68.5 & \vspace{2mm}0.030 &\\ [2.ex]
\hline
 \vspace{2mm}12  & \vspace{2mm}3.02 & \vspace{2mm}1.00 & \vspace{2mm}76.1 & \vspace{2mm}0.026 &\\ [2.ex]
\hline
 \vspace{2mm}13  & \vspace{2mm}2.48 & \vspace{2mm}1.25 & \vspace{2mm}89.6 & \vspace{2mm}0.021 &\\ [2.ex]
\hline
 \vspace{2mm}14  & \vspace{2mm}2.24 & \vspace{2mm}1.39 & \vspace{2mm}87.2 & \vspace{2mm}0.019 &\\ [2.ex]

\hline
\end{tabular}
\end{table}

\subsection{Simulated Galaxy Catalog}

Galaxies in the \refsim simulation were obtained through the SQL query available at the EAGLE website\footnote{\href{http://icc.dur.ac.uk/Eagle/database.php}{http://icc.dur.ac.uk/Eagle/database.php}} \citep{McAlpine2016}, where we retrieve their halo mass ($\mathrm{M_{200}}$), SFR and their rest frame U magnitude ($\mathrm{M_U}$, as an indicator of continuum emission). Since there is no information about the \lya\ luminosity of these galaxies, we do not know what fraction of them corresponds to the observed population of LAEs.  As a simple approximation, we only select galaxies with a quoted SFR (SFR$>0$).
Because star formation histories for galaxies with less than 100 stellar particles are affected by stochasticity, a fraction of the selected galaxies do not have an assigned $\mathrm{M_U}$ in the catalog. Ignoring those galaxies decreases the number density of galaxies below the observed number density. To keep a sufficient number of simulated galaxies in our sample, we manually assign a value of $\mathrm{M_U}$ for these galaxies based on the instantaneous relation between $\mathrm{M_U}$ and SFR in the catalog. Deviations from this relation are expected to be stochastic and therefore to cancel out on average in our analysis based on stacking.

Similarly as with the MUSE cubes, subcubes were extracted around the 3-D positions of the selected galaxies.

\subsection{Uncertainties in the predicted $\mathrm{f_{LLS}}$ from simulations}

The uncertainties involved in our predicted gas density distribution, our assumption of self-shielded clouds as the only sources of \lya\ photons, and the fact that the galaxies selected in the simulations may not correspond to the observed LAEs will all have an effect on the predicted \flls\ radial profile. These effects are relevant for our estimate of \G\ as discussed below. 

\subsubsection{Column density distribution}\label{coldensdist}

\begin{figure}
\begin{center}
\includegraphics[trim={0.2cm 0cm 0cm 0cm},clip,width=.43\textwidth]{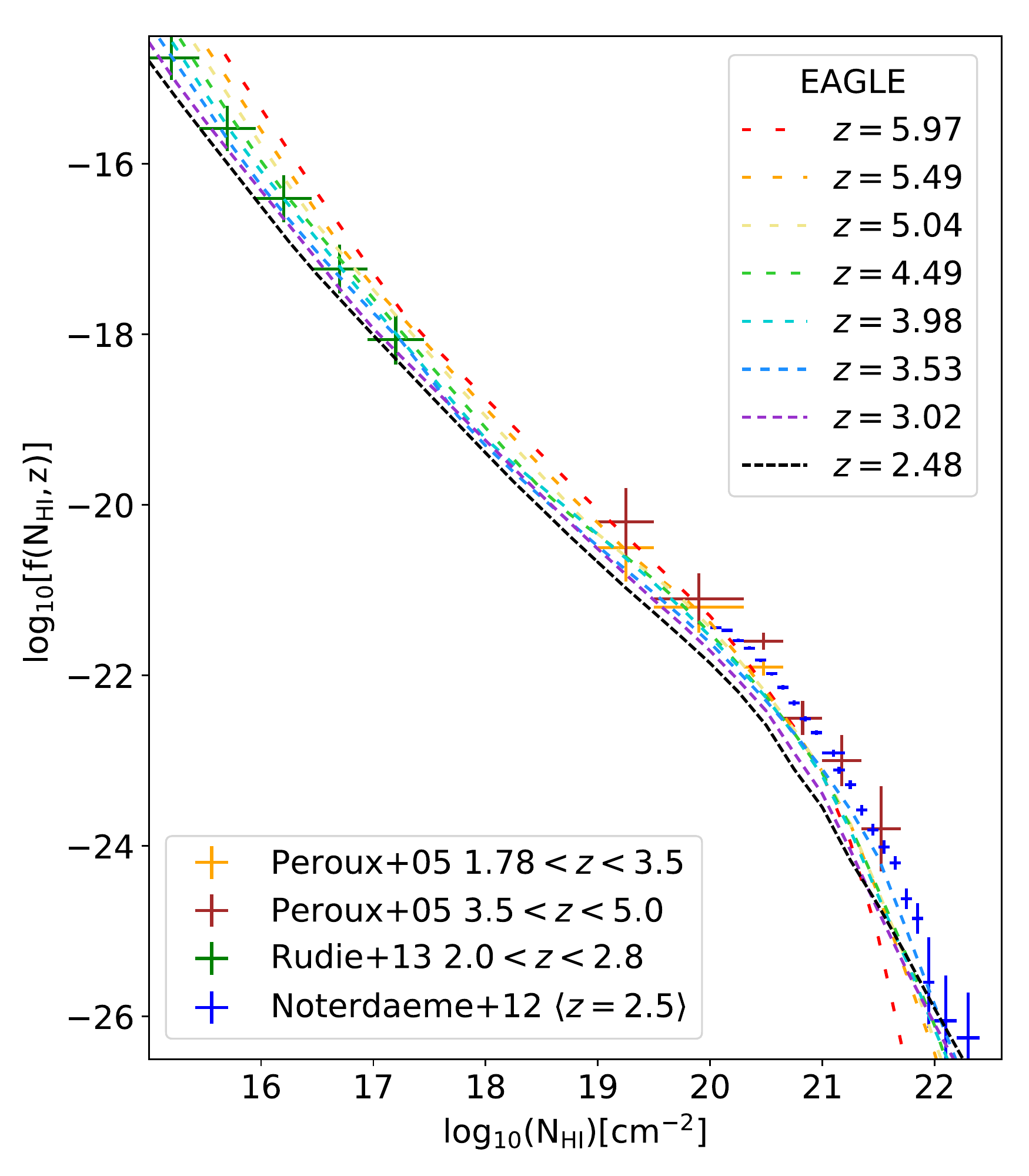}
  \caption{HI column density distribution function for the EAGLE mock cubes from the full simulation box. For comparison, observational data from absorption line studies is shown \citep{Peroux2005,Noterdaeme2012,Rudie2013}.}  \label{cddf}  
\end{center}
\end{figure} 

\begin{figure}
\begin{center}
\includegraphics[trim={0cm 0cm 0cm 0cm},clip,width=.5\textwidth]{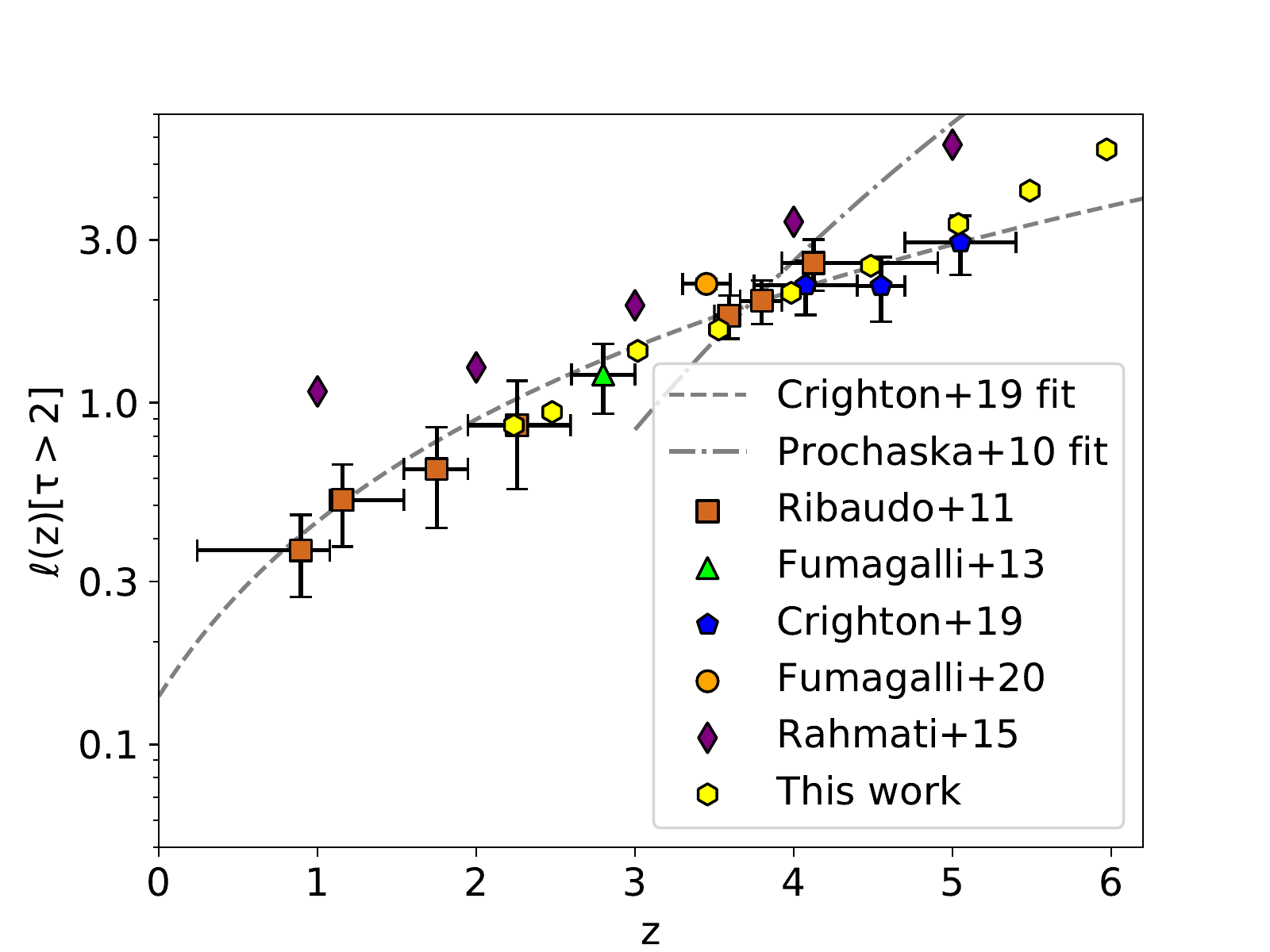}
  \caption{Number of LLSs per unit redshift obtained from the EAGLE \NHI\ mock cubes from \citet{Rahmati2015} and our work compared to observational constraints \citep{Prochaska2010,Ribaudo2011,Fumagalli2013,Crighton2019,Fumagalli2020}. We define a LLS as having an optical depth at the Lyman limit above 2, which corresponds to an HI column density of $10^{17.5}\,\mathrm{cm^{-2}}$.}    \label{dndz}
\end{center}
\end{figure}

To quantify the accuracy of our estimated column densities at different redshifts from EAGLE with respect to observations we calculate the column density distribution function (CDDF), or equivalently the number of absorbers per unit column density ($d\,\mathrm{N_{HI}}$) per unit absorption length ($d\,X=d\,z (H_0/H_z)(1+z)^2$). As shown in Fig. \ref{cddf} our results agree well with the observational data, although we are below the observational results at higher column densities. The small disagreement at large NHI can be explained by the small number of massive halos for the chosen simulation box and the decrease in resolution when converting particles from EAGLE into a grid, which can smooth-out the highest densities. Nonetheless, a remarkable agreement is found when calculating $l(z)$ (Fig. \ref{dndz}), the number of absorbers above the Lyman limit threshold per unit redshift. Our values, compared the latest observational results \citep[][]{Crighton2019}, are in much better agreement with respect to the estimated $l(z)$ from the CDDF of \citet[][]{Rahmati2015}\footnote{\textit{Ref-L100N1504}, adjusted to the \citet{Bennett2014} cosmology.}, which may be partly coincidental or partly due to the higher resolution of our chosen simulation box and/or to the different adopted UVB model. We stress however, that the detailed physical origin of the particular LLSs covering fraction used here are not important as long as they can be considered a good approximation of the real data. 

\subsubsection{Self-shielding transition}\label{self}

\begin{figure}
\centering
\includegraphics[trim={0.7cm 0cm 2cm 0cm},clip,width=.5\textwidth]
{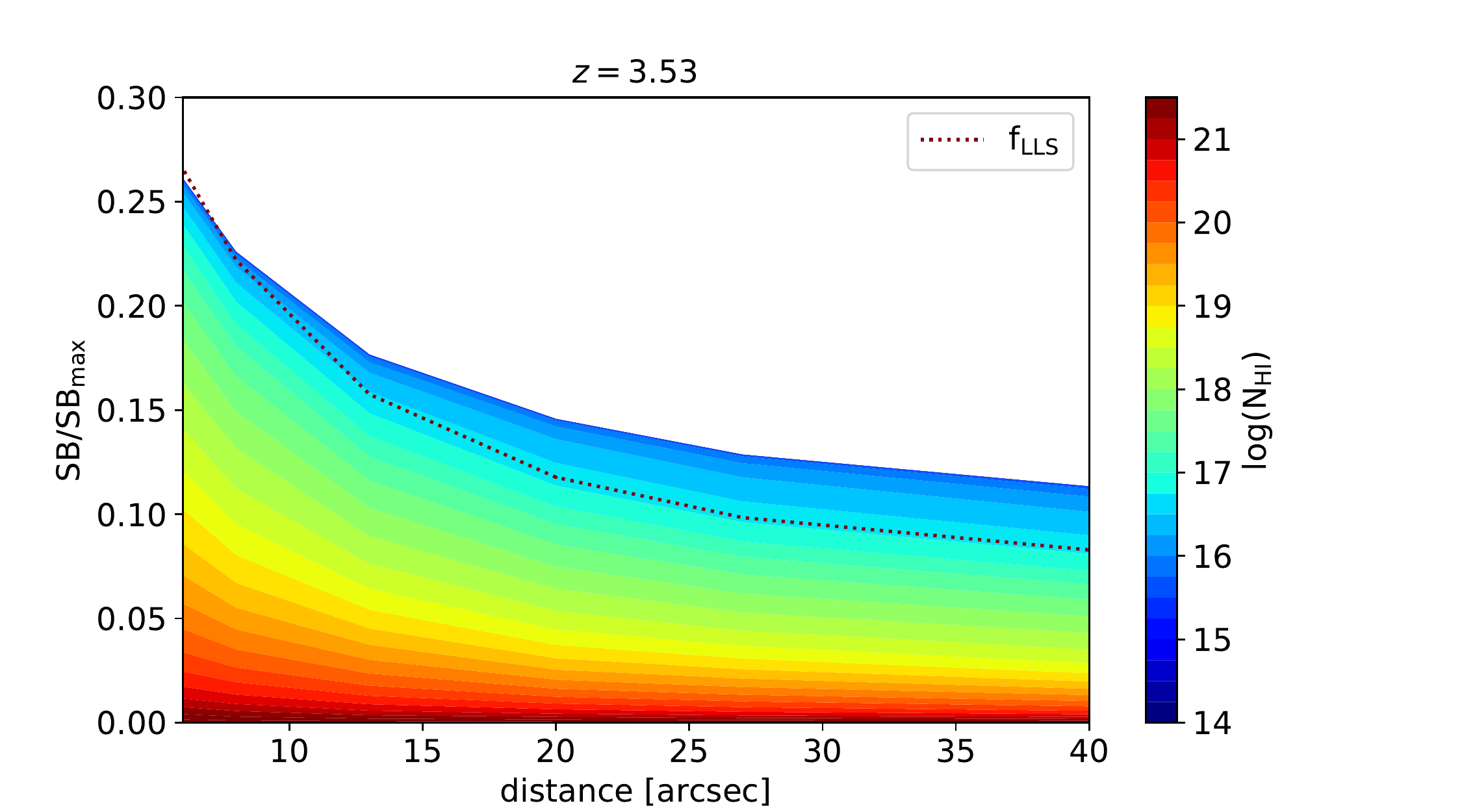}
   \caption{EAGLE \lya\ SB profile expectations at redshift 3.5, color coded by the differential contribution of different \NHI\ column densities, and normalized by the maximum expected emission from the UVB fluorescence. The \lya\ SB was calculated using the \citet{Cantalupo2005} conversion between SB and \NHI, multiplied by the corresponding \cf\ of each \NHI\ bin. The dashed line represents the \cf\ of LLSs (i.e. \NHI$>10^{17.5}\,\mathrm{cm^{-2}}$). If column densities above the LLS threshold would emit 100\% of the \lya\ SB from fluorescence and everything below that threshold did not emit any \lya, then the expected SB would match the dashed curve ($\mathrm{f_{LLS}}$). } 
   \label{SBE}
\end{figure}

\begin{figure}
\centering
\includegraphics[width=.5\textwidth]{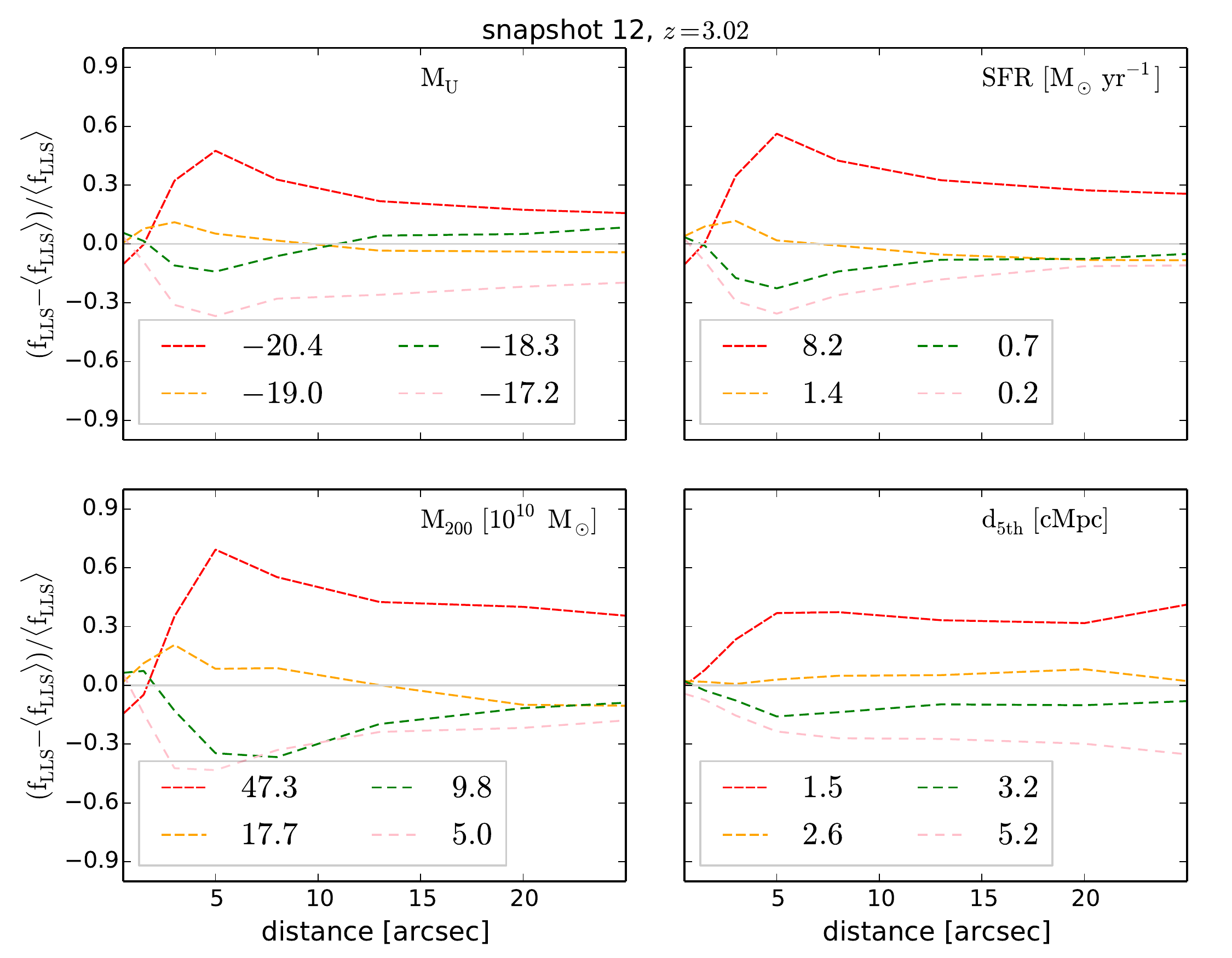}
   \caption{Differential \flls\ at redshift 3.02 for quartiles of different properties of the central simulated galaxies. The quoted number represents the mean of each quartile.}    \label{lls12}
\end{figure}

\begin{figure}
\centering
\includegraphics[trim={.6cm 0cm 0.4cm 0.3cm},clip,width=.5\textwidth]{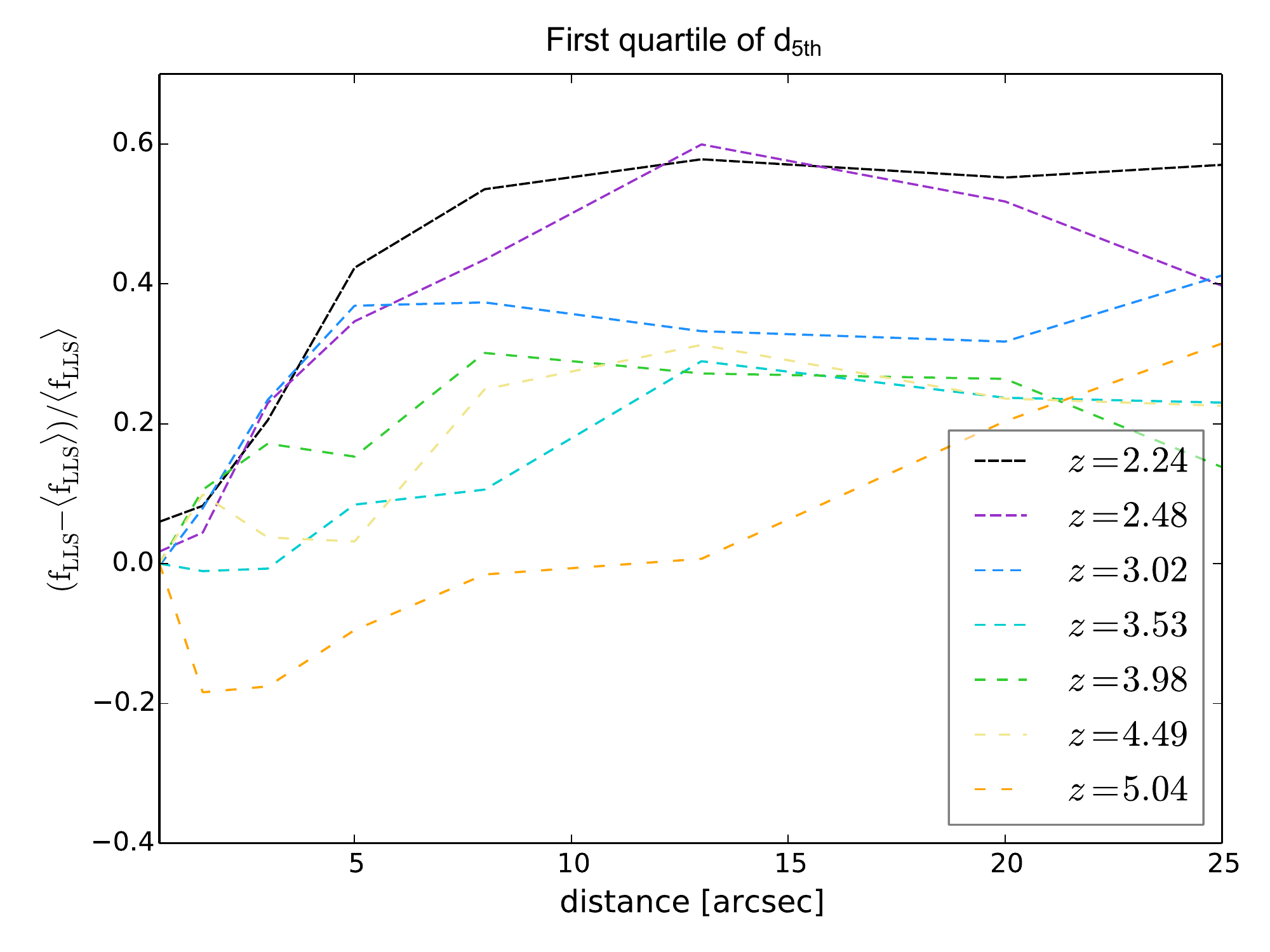}
   \caption{Redshift evolution of the differential \flls\ for the first quartile of $\mathrm{d_{5th}}$ (distance of the central galaxy to its fifth neighbor).}    \label{n5d}
\end{figure}

As described in section \ref{sec:uvb}, our \G\ prediction from equation \ref{maxSB} accounts for the maximum SB expected for a single self-shielded cloud, which happens above \NHI\ $\approx10^{18}\mathrm{cm^2}$. In practice, part of the \lya\ emission will come from optically thin clouds (\NHI\ $<10^{17}\mathrm{cm^2}$) and especially between the transition between optically thin and optically thick clouds, where there is no observational constraints to date. Moreover, in the case of extended emission, an accurate estimation will require to consider the fraction of the observed region that is covered by self-shielded clouds. We chose to account for both effects in $\mathrm{f_{LLS}}$, which we assume to be the \cf\ of LLSs.

To estimate how accurate the approximation of a sharp transition between self-shielded and not self-shielded gas is, we convert a stacked $\mathrm{N_{HI}}$ profile (before converting to \flls) around galaxies at $z=3.53$ into an expected SB (Fig. \ref{SBE}) using a simple fitting function between \NHI\ and SB from the output of the simulations presented in \citet[][Fig. 6]{Cantalupo2005}. If column densities above the LLS threshold emit 100\% of the maximum expected \lya\ SB from fluorescence and everything below that threshold does not emit any \lya, then the expected SB would match the dashed curve shown in Fig. \ref{SBE}.
 As we will see in section \ref{sec:results}, some of our observational results are upper limits on the expected SB. Given that our mock cubes slightly underestimate the abundance of high $\mathrm{N_{HI}}$ gas with respect to observational constraints, we choose not to correct for this effect.

\subsubsection{Contribution of local ionizing sources in simulations}\label{sim:localUV}

One of the main limitations of our mock cubes is that the impact of local ionizing sources on the \NHI\ distribution and \lya\ emission is ignored. This can have two opposite consequences for the expected SB: an increase in the local ionizing radiation (increasing the rate of \lya\ photons) and a decrease in the neutral hydrogen density (decreasing the rate of \lya\ photons) due to that same radiation ionizing the gas. The relative importance of these effects will depend on the ionizing escape fraction from galaxies and the morphology of the surrounding gas, which are difficult to model. Given that we expect the contribution of local sources to be larger the closer we are to the central galaxies in our stacks, our estimated covering fractions will be less reliable at distances typical of the CGM, which makes any prediction at those scales uncertain. In order to have a better comparison, a full radiative transfer simulation with sufficiently large resolution for both the galaxies and their CGM would be necessary. In absence of these models, the currently simulated covering fraction could be considered as upper limits. We note however, that numerical resolution could also have an important effect and some recent models suggest that increasing the resolution of the simulations should increase the covering fraction of LLSs, possibly balancing to some degree the lack of local ionizing sources.

\subsubsection{Dependence of the covering fraction on galaxy properties}\label{sec:galprops}

In order to obtain our simulated \flls\ radial profiles, we include all galaxies in the simulation with a measured SFR, which may in reality not correspond to our observed LAEs. Since there is no clear way to separate LAEs from non-LAEs in the simulations,
we investigate the relation between the properties of the central galaxies and the radial \flls\ profiles. In Figure \ref{lls12} (more in Fig. \ref{LLSsnaps}) we plot the relative difference between the average profile and the profile for different bins of rest-frame U magnitude ($\mathrm{M_U}$), SFR, halo mass ($\mathrm{M_{200}}$) and the distance of the central galaxy to its fifth neighbour ($\mathrm{d_{5th}}$), for $z=3.02$. 

In the case of $\mathrm{d_{5th}}$ (Fig. \ref{n5d}), the enhancement becomes more relevant above 20" for $z=3$ and $z=2.2$. These results suggest that the IGM among more clustered galaxies is consistently denser up to very large distances. Since $\mathrm{d_{5th}}$ is a much easier quantity to estimate observationally than the others -it only requires the position and redshift of the galaxies for a homogeneously sampled catalog-, it is potentially an excellent tool for surveys covering a large FoV.


\section{Results}\label{sec:results}

In this section we describe the results obtained on the \lya\ stacking around MUSE galaxies, from which we obtain the observational constraints on \G\ assuming either 100\% covering fraction of LLS, or the \cf\ of LLS on mock cubes produced from the EAGLE simulation. Moreover, we constrain \flls\ around LAEs based on our observational results and assuming the latest UVB models.
 
\subsection{MUSE stacked spectra}

\begin{figure*}
\centering
\includegraphics[trim={.5cm 1.2cm 0.4cm 0.3cm},clip,width=.5\textwidth]
{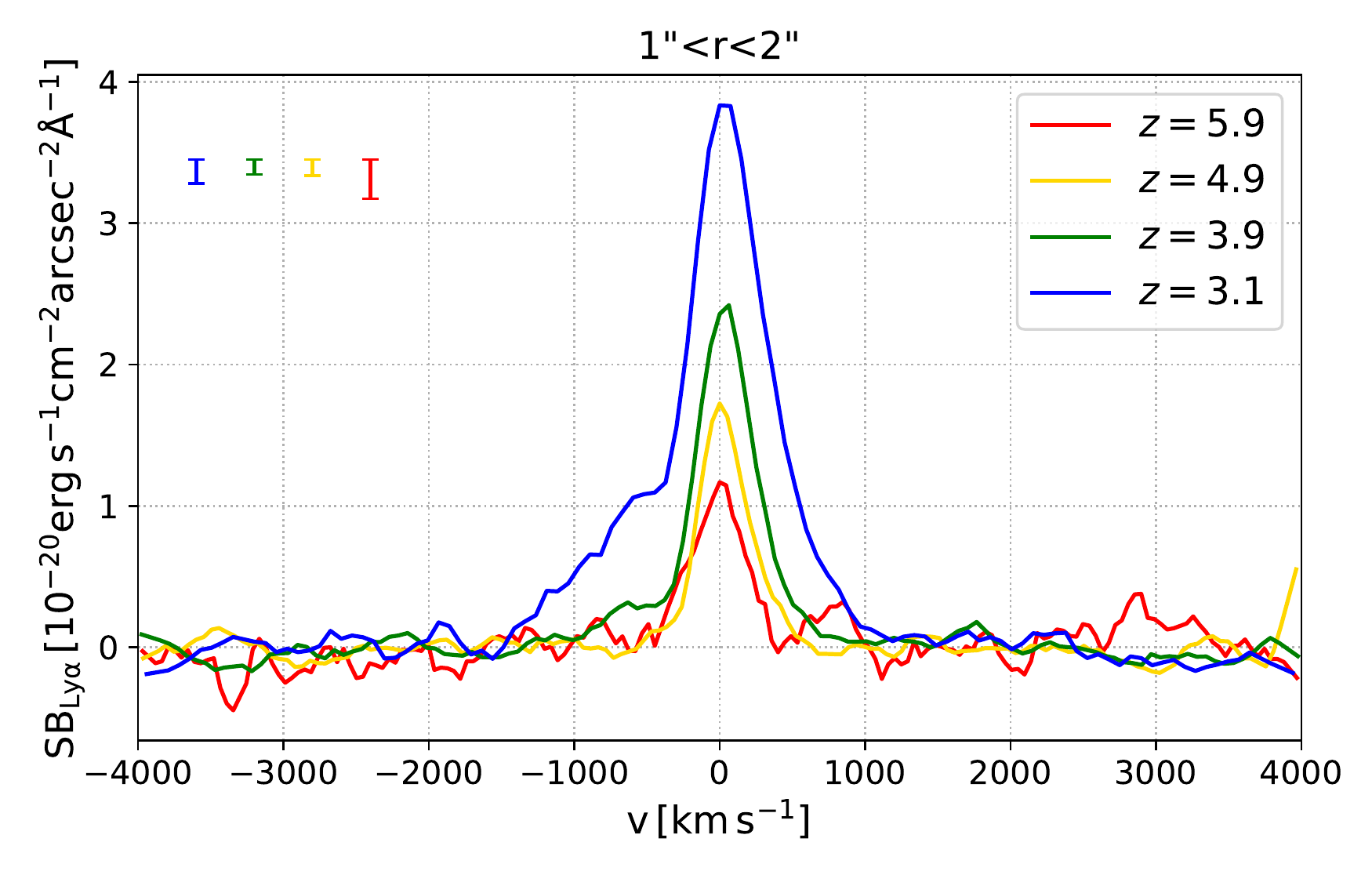}
\includegraphics[trim={1.2cm 1.2cm 0.4cm 0.3cm},clip,width=.48\textwidth]
{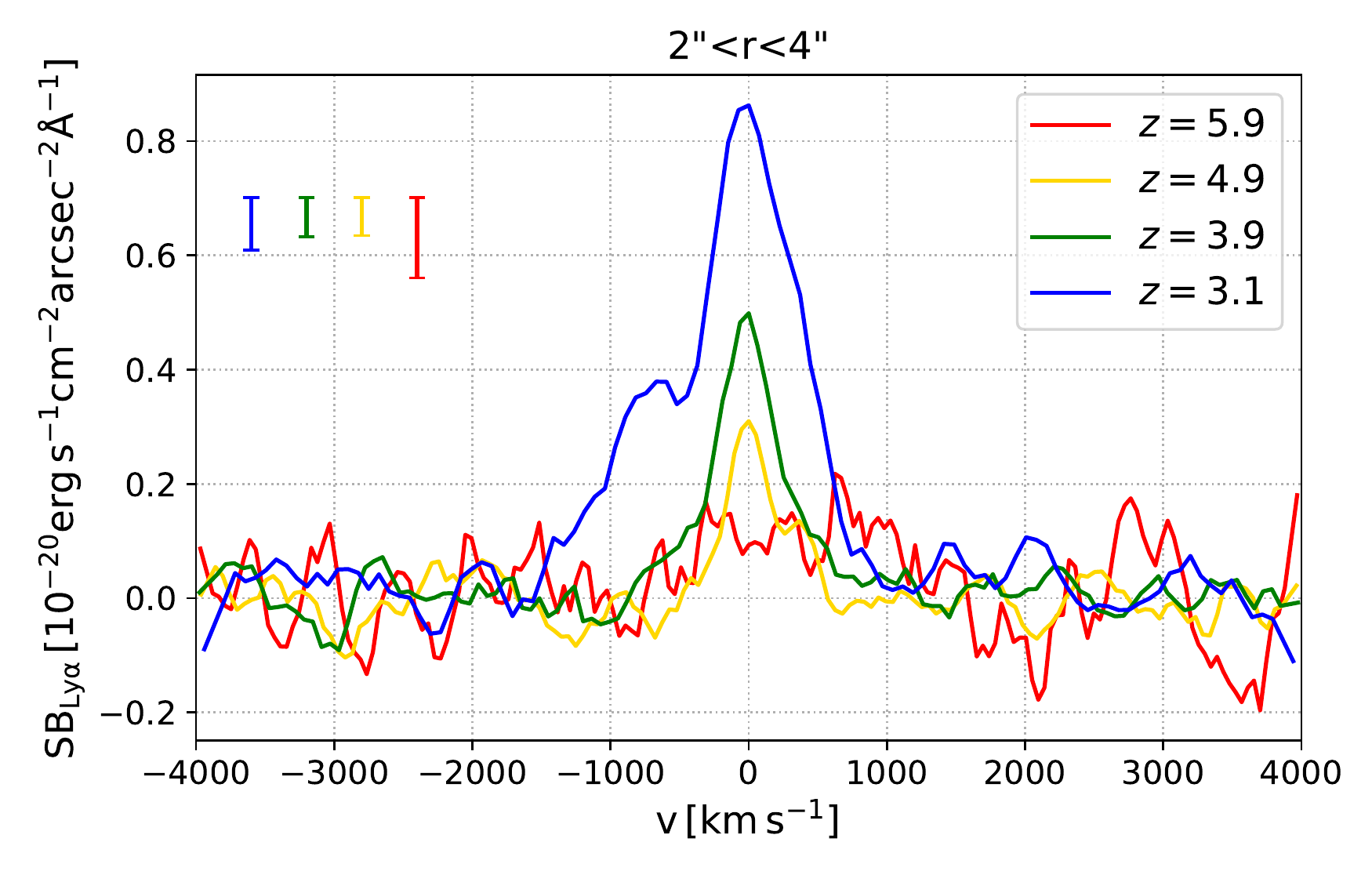}
\includegraphics[trim={.5cm 1.2cm 0.cm 0.3cm},clip,width=.5\textwidth]
{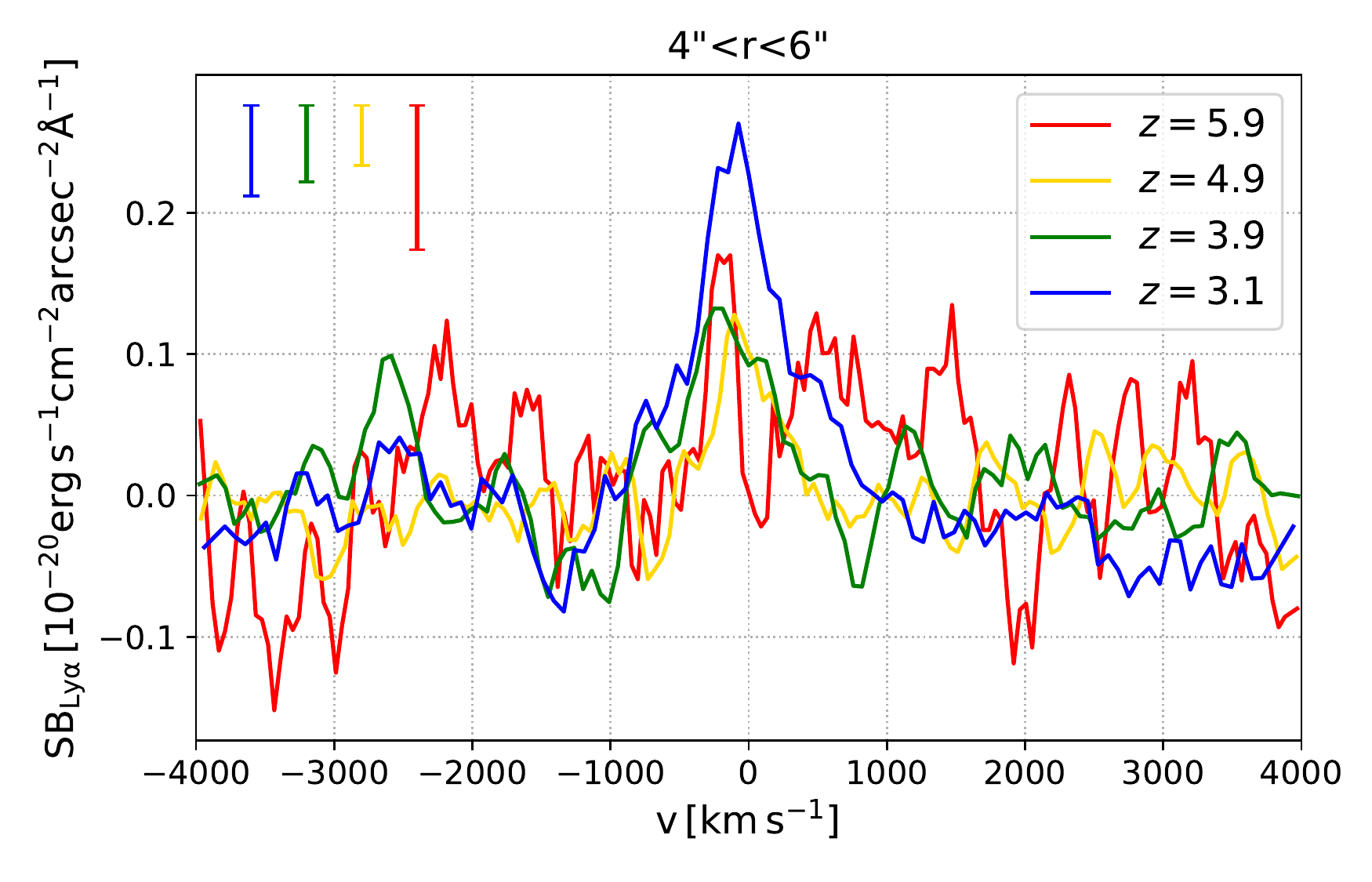}
\includegraphics[trim={1.2cm 1.2cm 0.cm 0.3cm},clip,width=.48\textwidth]
{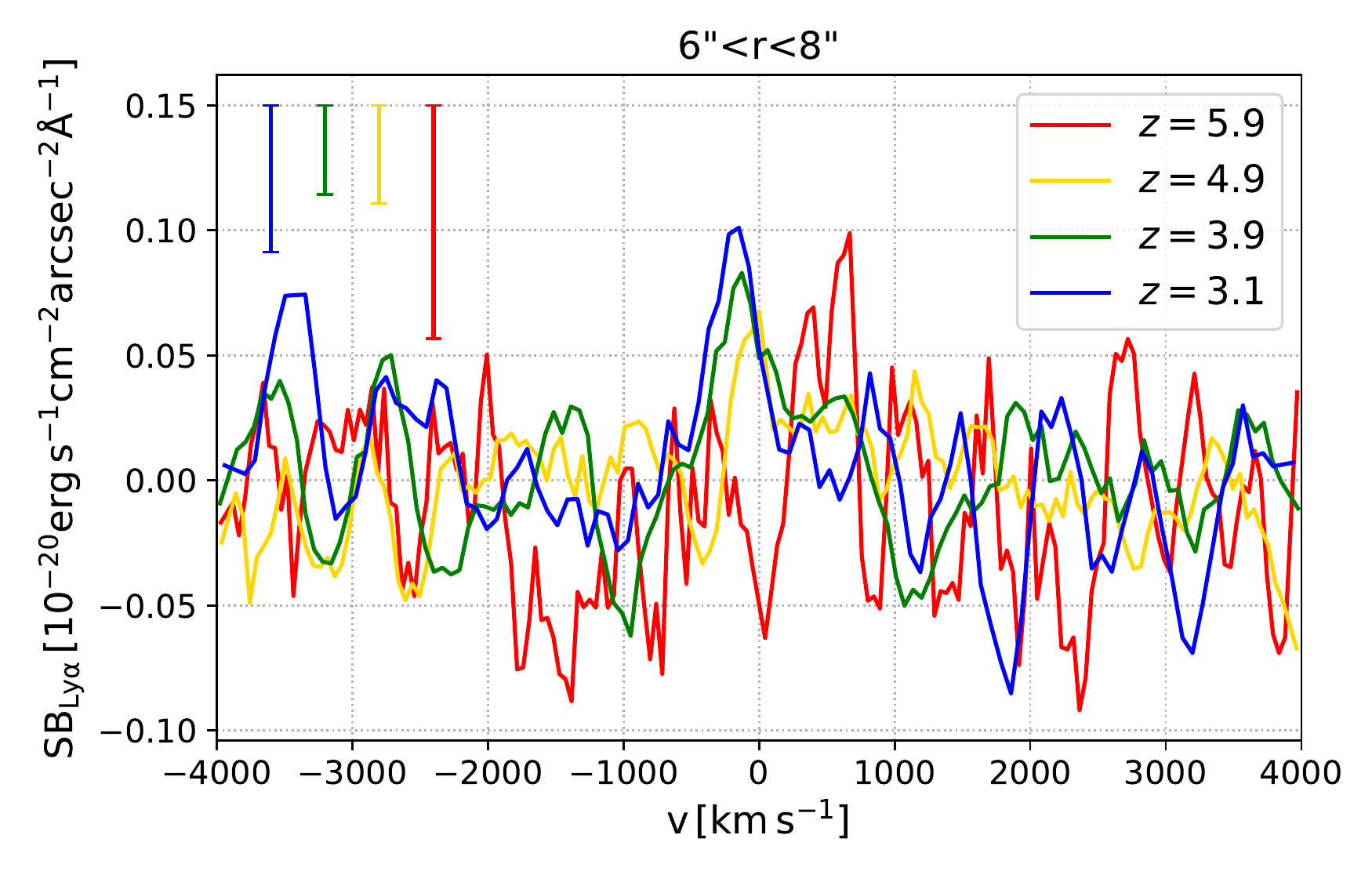}
\includegraphics[trim={.5cm .6cm 0.cm 0.3cm},clip,width=.5\textwidth]
{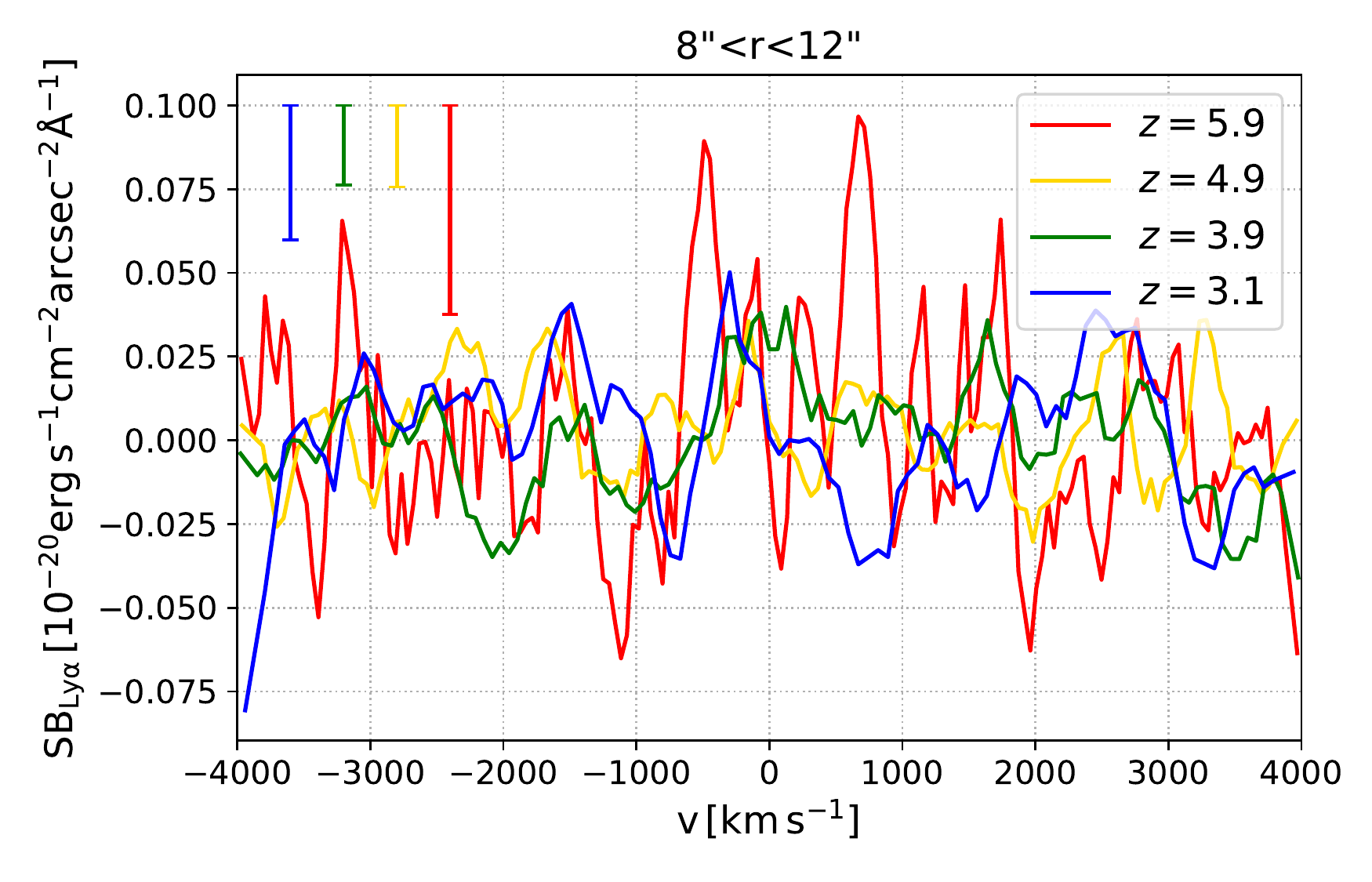}
\includegraphics[trim={1.2cm .6cm 0.cm 0.3cm},clip,width=.48\textwidth]
{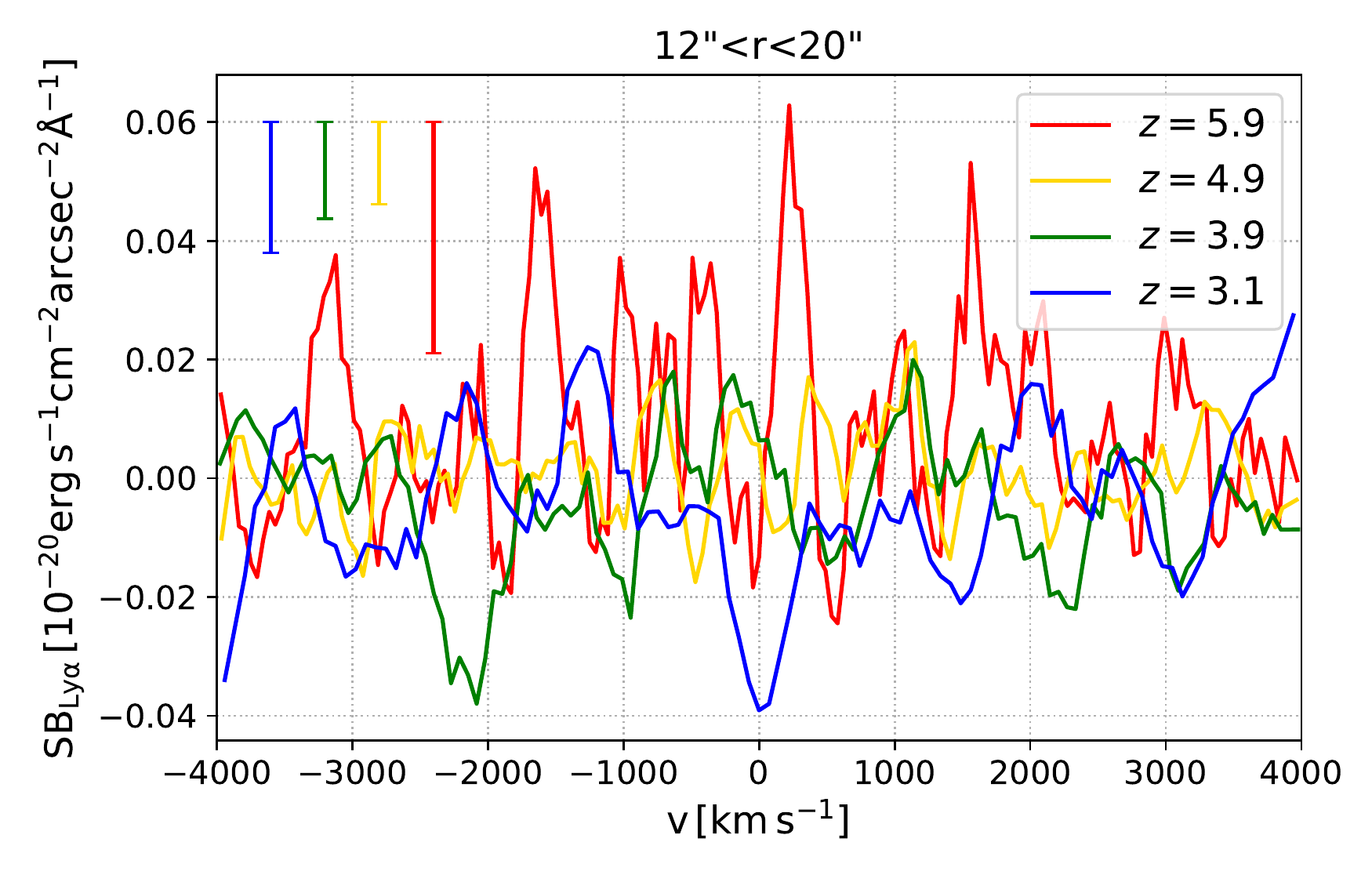}
   \caption{Stacked spectra at different radial annuli around LAEs. The center is chosen as the 3-d peak of the \lya\ emission.  The vertical lines are the 1$\sigma$ SB limit per redshift bin estimated on all layers excluding those within $\pm2000\,\mathrm{km\,s^{-1}}$ from the \lya\ peak. 
   The spectra have been smoothed for visualization purposes. Note the different vertical axis scales.
   }    \label{specr}
\end{figure*}

\begin{figure*}
\centering
\includegraphics[trim={.5cm 1.8cm 0.4cm 0.3cm},clip,width=.5\textwidth]
{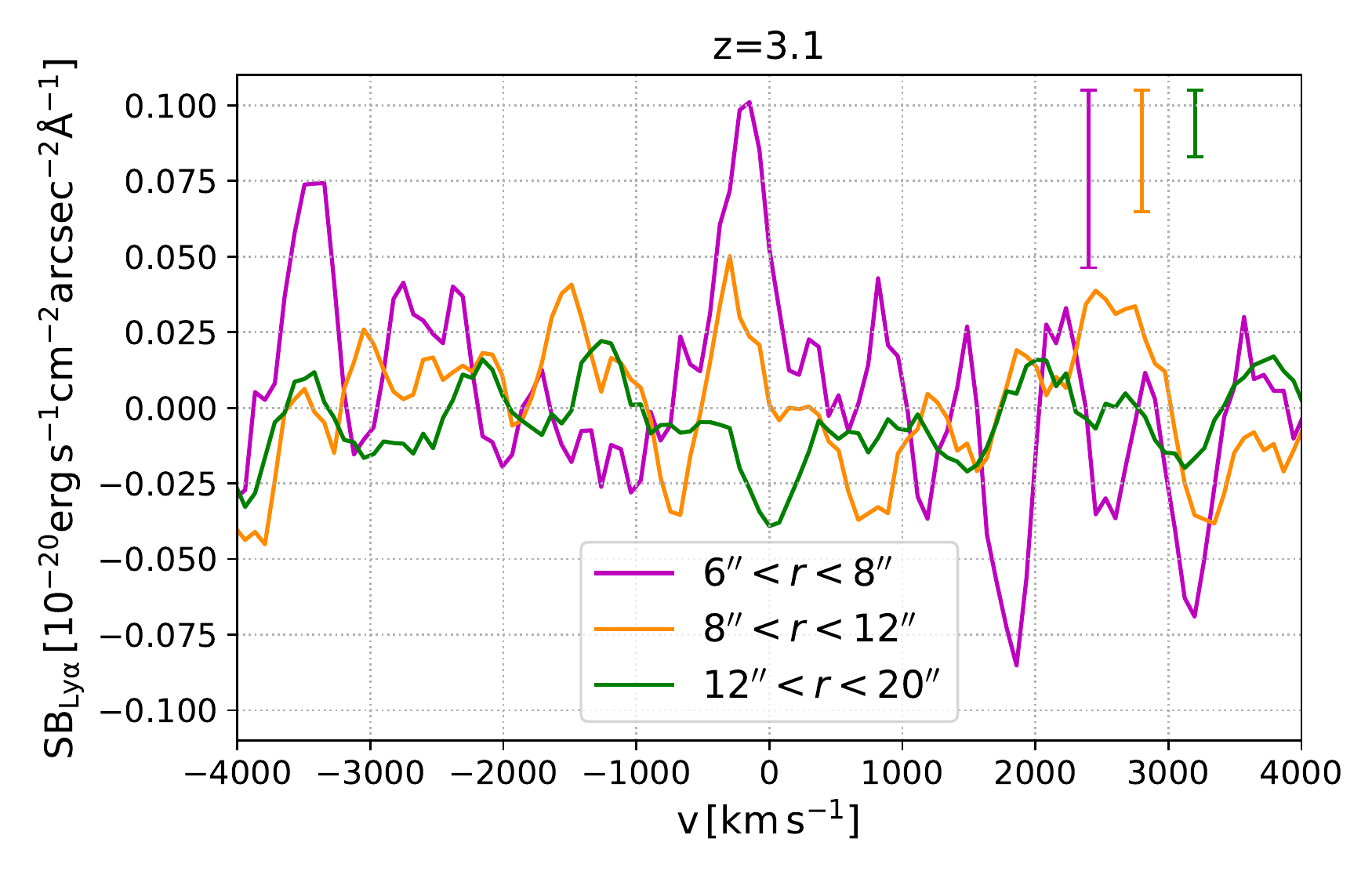}
\includegraphics[trim={1.2cm 1.8cm 0.4cm 0.3cm},clip,width=.48\textwidth]
{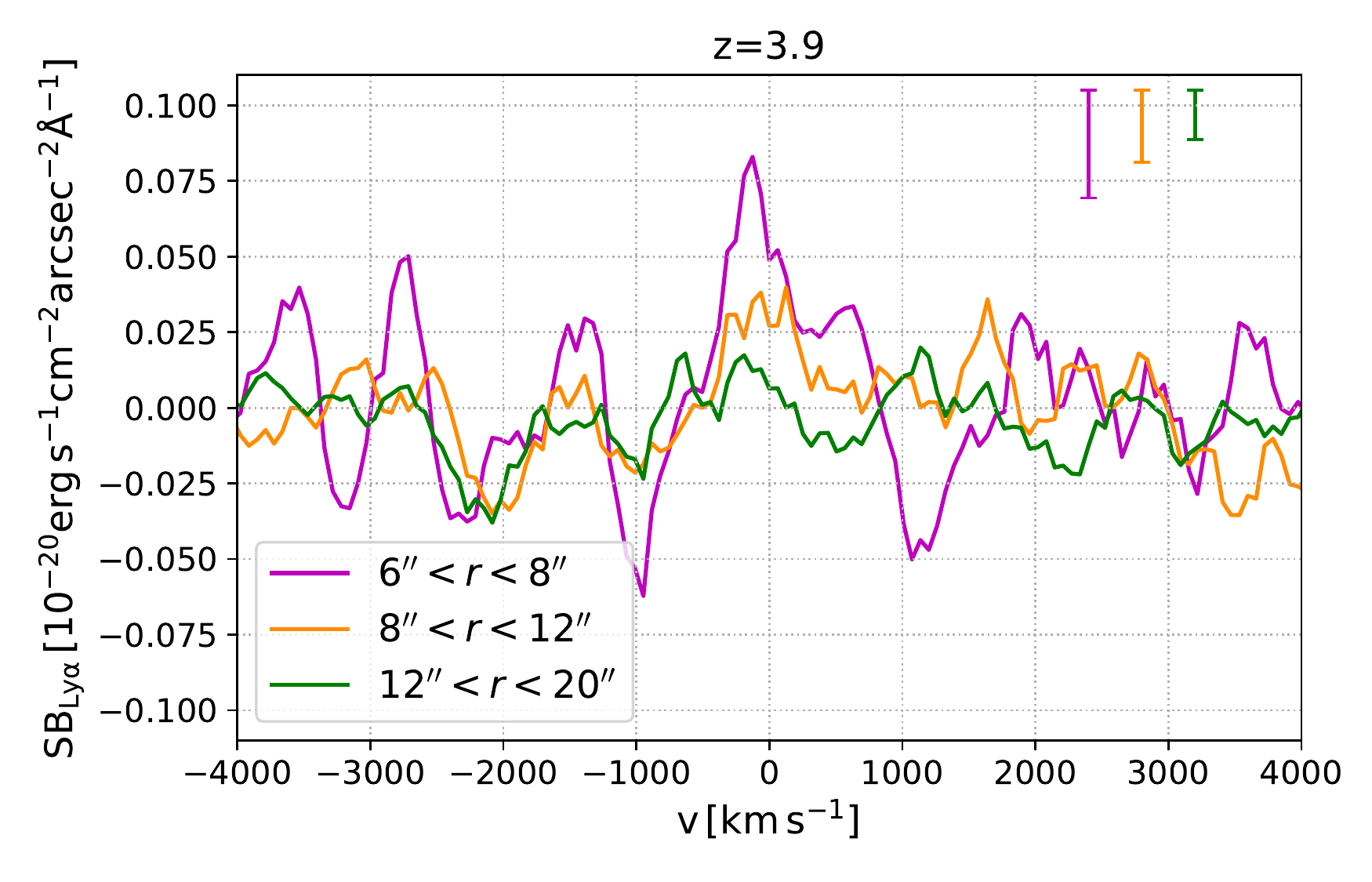}
\includegraphics[trim={.5cm .0cm 0.4cm 0.cm},clip,width=.5\textwidth]
{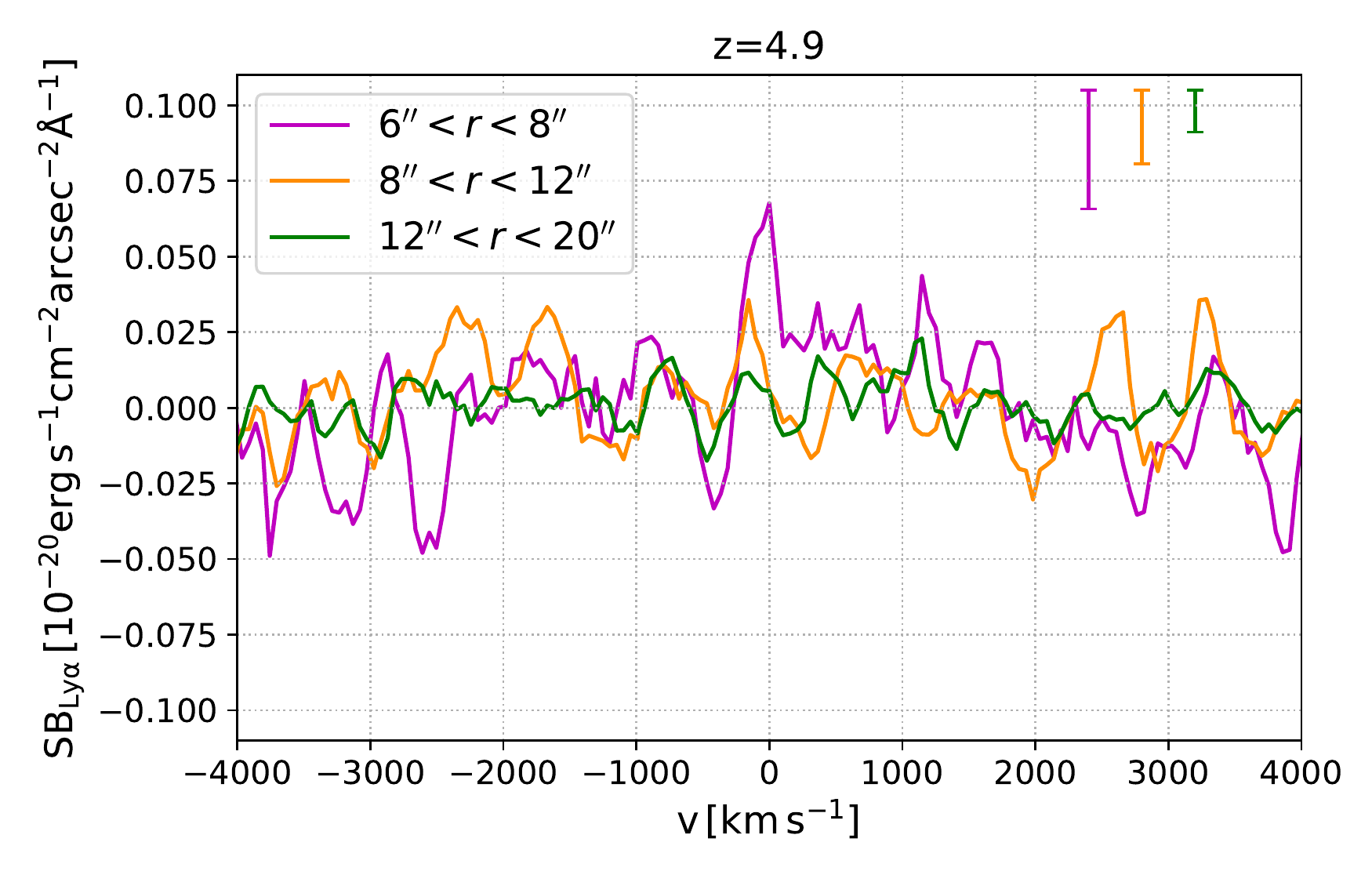}
\includegraphics[trim={1.2cm .0cm 0.4cm 0.cm},clip,width=.48\textwidth]
{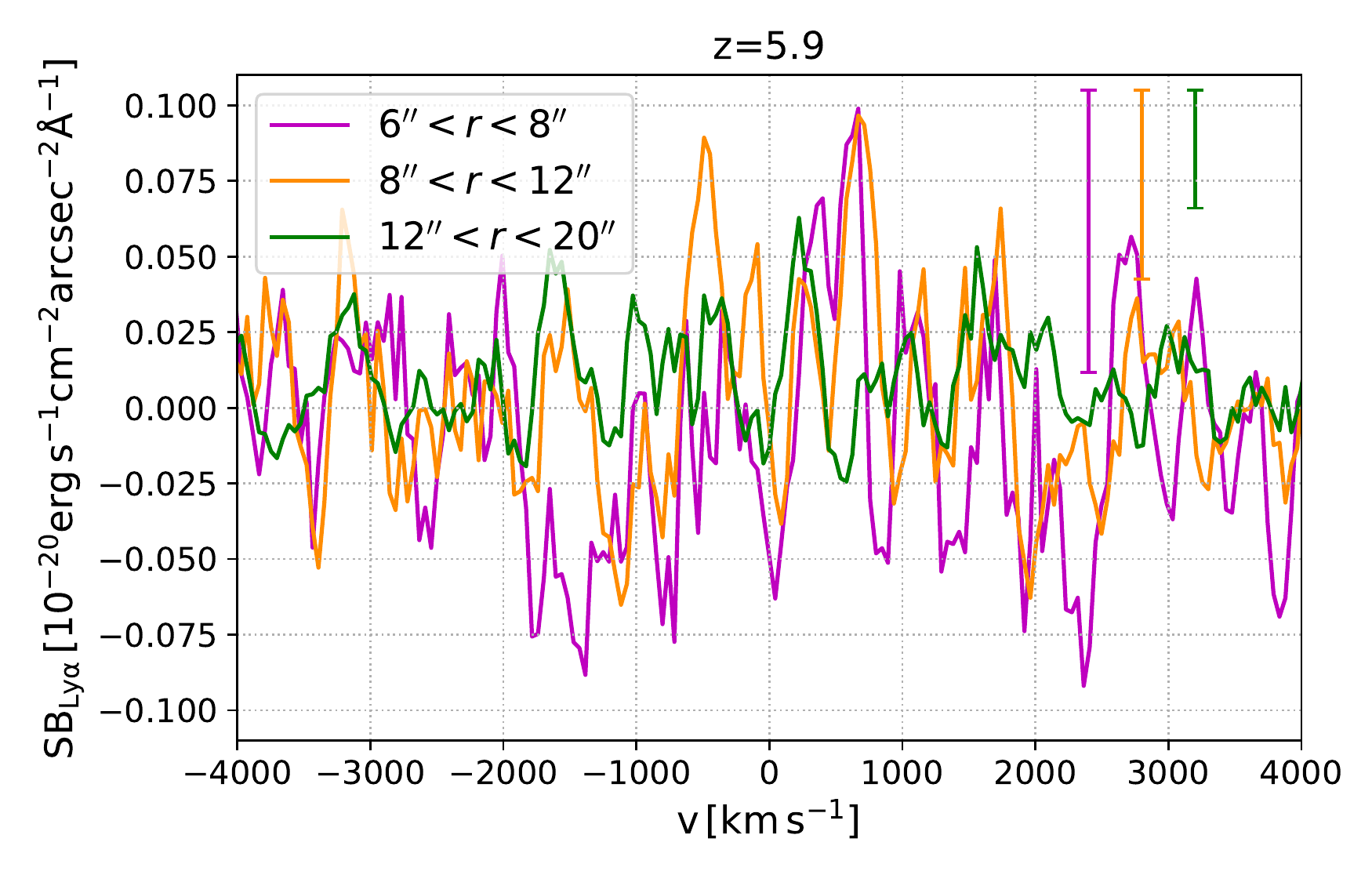}
   \caption{Same as Figure 2 but showing the different radial bins at each redshift in order to better visualise the significance or absence of a possible detection individually for each redshift bin. The radial bins below $6"$, which are very likely associated with the galaxy haloes themselves, are not shown here for clarity .}\label{specz}
\end{figure*}  

Figures \ref{specr} and \ref{specz} show the stacked spectra of the selected redshift bins in different radial annuli, centered on the spatial peak of the \lya\ emission. 

There is a clear detection of \lya\ emission up to 8" for all redshift bins except the one at $5.9$. The non detection at $z=5.9$ could be explained by the presence of more prominent skylines and the lower number of galaxies stacked which increase the overall noise.  
For the lower redshift bins, at distances above 8" we reached a $2\sigma$ SB limit of about $4\times 10^{-22}\,\SBcgs$, although there is no detection of extended emission. The SB limit was evaluated as the standard deviation among the same chosen aperture on layers $\pm 2000\,\mathrm{km\,s^{-1}}$ away from the central layer to avoid overestimating the noise due to the central \lya\ emission.

Between 6" to 8" for the first two redshift bins, there is an indication that the peak of the \lya\ emission is blueshifted with respect of the peak of \lya\ emission of the central galaxy, this corresponds to about $3.75\,\mathrm{\text{\AA}}$ ($\approx-200\,\mathrm{km\,s^{-1}}$). 
Previous studies have found that \lya\ emission lines of LAEs are systematically redshifted with respect to the systemic redshifts of galaxies traced by additional lines \citep{Shapley2003,McLinden2011,Rakic2011,Song2014,Hashimoto2015}. Most recently, \citet{Muzahid2019} found an offset of $\approx180\,\mathrm{km\,s^{-1}}$ in a sample of MUSE detected LAEs by stacking their CGM absorption lines in background quasars. Our results seem to agree with previous studies and suggest that CGM \lya\ emission is a better tracer of the true systemic redshift of LAEs. 

\subsection{Surface brightness profiles}

\begin{figure}
\centering
\includegraphics[trim={.5cm .0cm 0.4cm 0.3cm},clip,width=.5\textwidth]
{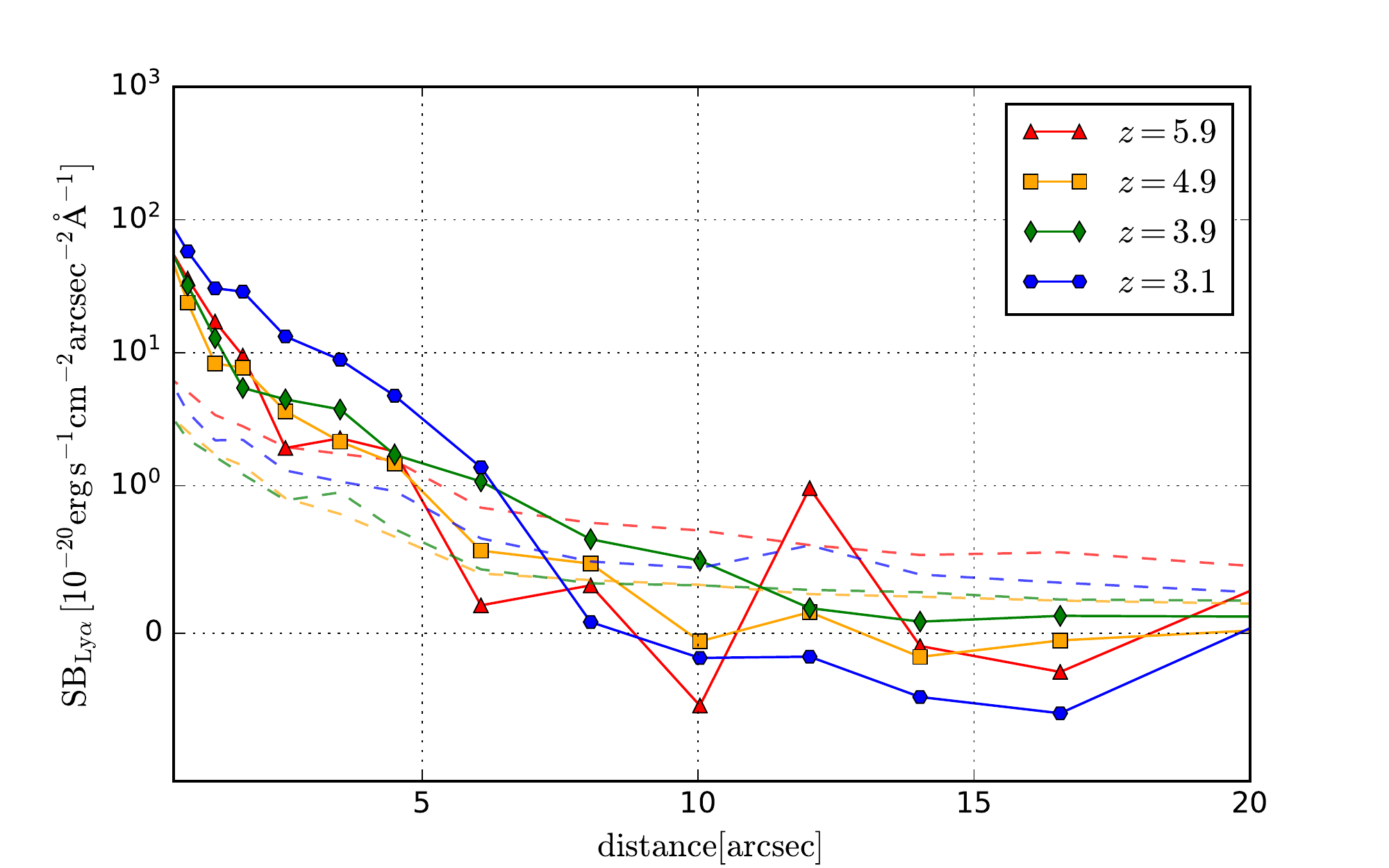}
\includegraphics[trim={.5cm .0cm 0.4cm 1cm},clip,width=.5\textwidth]
{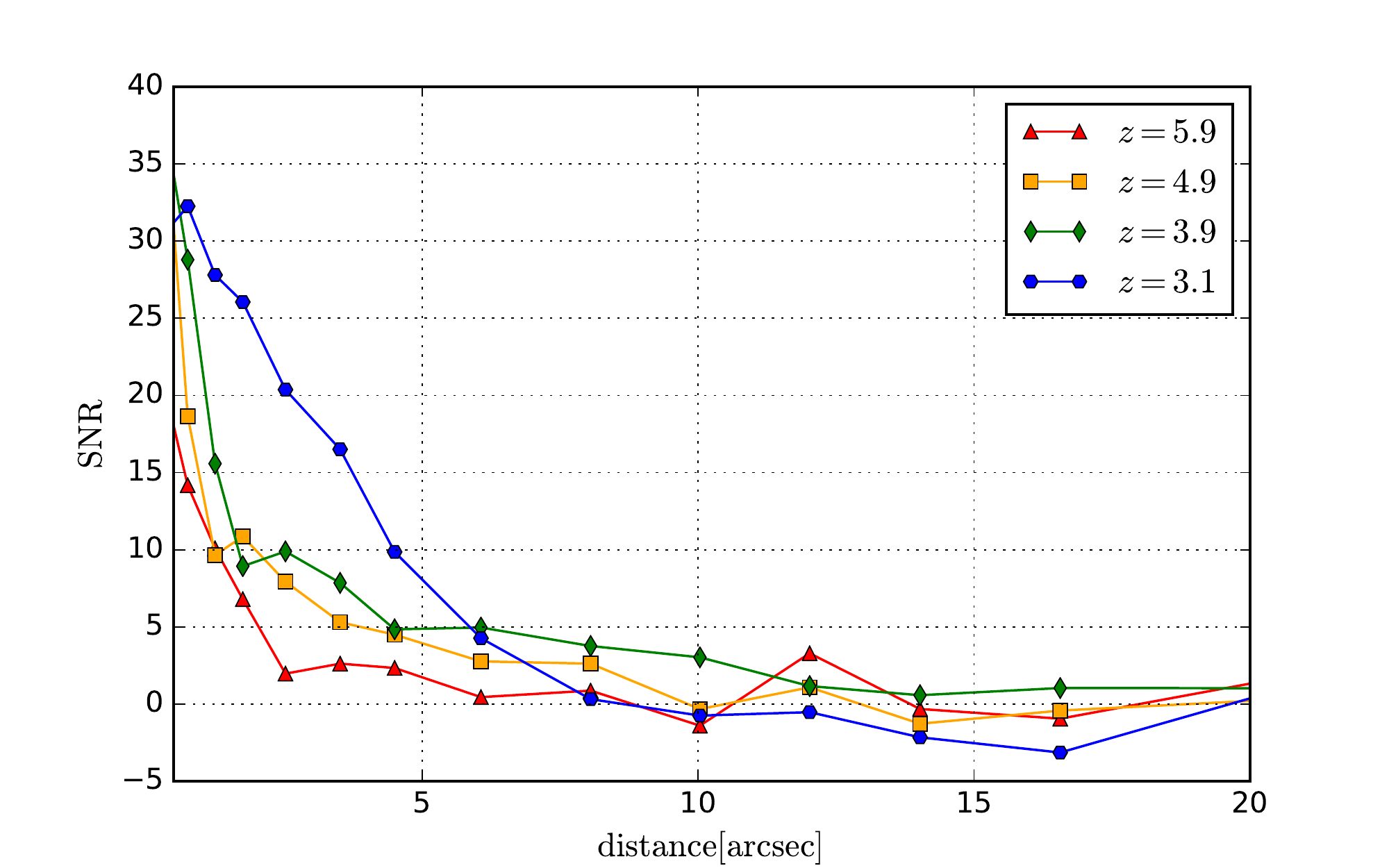}
   \caption{Top: Stacked SB profiles in function of distance from the galaxies for the different redshift bins probed in this study. The coloured dashed lines correspond to the $2\sigma$ noise levels. Bottom: Stacked signal-to-noise profiles.} \label{SBSNR}   
\end{figure} 

Figure \ref{SBSNR} shows the SB and signal-to-noise ratio (SNR) radial profiles for the different redshift bins, with a wavelength width of $18.75\,\mathrm{\text{\AA}}$. We choose to recenter our stacks by the observed shift of $3.75\,\mathrm{\text{\AA}}$ with respect to the peak of the \lya\ emission of the central galaxies. The coloured dashed lines correspond to the $2\sigma$ noise levels. \lya\ emission is detected (above 2$\sigma$) for $z=3.9$ out to a maximum 10" ($\approx70\,\mathrm{kpc}$), slightly above the detection levels in the stacked spectra due to integration over a larger wavelength width. The profiles seem in agreement with the work of \citet{Wisotzki2016}, but reaching a higher SNR at large radii.

From the works of \citet{Wisotzki2016} and \citet{Leclercq2017}, based on a two-component model for the radial profiles of the galaxy and halo, we know that the characteristic scale length for \lya\ halos of LAEs is about 4.5 kpc, with some of the halo's scale lengths extending up to 20 kpc (below 3"). By focusing on the regions beyond 8", i.e. beyond about 63 kpc (at $z\approx3$), we are confident to be enough far away from the central regions of the halos which could be affected by processes related to the galaxies themselves. 
Regarding the possible contribution of the local radiation field, we cannot exclude that some of our galaxies could be bright enough
(especially in the case in which they could harbor an AGN or if they have for some reason a large escape fraction of ionizing photons) to contribute to the \G\ up to that distance. However, we think that by performing a stacking analysis these cases should not significantly affect our measurement, unless they are common. In the latter case, our estimates of \G\ should be considered as upper limits. 

\subsection{\G\ observational limits assuming a LLS \cf}

If we assume that at a distance larger than 8" away from the observed galaxies all of the \lya\ emission originates from \lya\ fluorescence of optically thick HI clouds, that the ionizing light comes exclusively from the UVB and that we know the covering fraction of LLSs in the selected area, then we can make a first evaluation of the value of \G\ directly from the observations. 

\begin{table*}
\centering
\caption{Summary of observational results presented in this work. Columns include the selected redshift range, mean redshift, number of LAEs above confidence 2 in the MUSE catalog, expected SB for the HM12 model in $10^{-20}$ $\SBcgs\ $ and the measured SB on the annular region between 8" and 20" away from galaxies, the predicted \G\ in $10^{-12}\,\mathrm{s}^{-1}$ for a \cf\ $\mathrm{f_{LLS}}=1$, the last two columns are the predicted f$_{\mathrm{LLS}}$ on the same region obtained with equation \ref{gcalc}, using the HM12 and \citet[][HM01]{Haardt2001} UVB constraints. The quoted upper limits are 2$\sigma$ above the mean.
}
\label{gobs}
\begin{tabular}{| >{\centering}m{1.3cm} | >{\centering}m{1.2cm} | >{\centering}m{1.8cm} | >{\centering}m{1.5cm} | >{\centering}m{2.2cm}@{} | >{\centering}m{2.2cm}@{} | >{\centering}m{2.2cm}@{} | >{\centering}m{2.2cm}@{}    m{0cm}@{} |}
\hline
\vspace{2mm}$z$ range & 
\vspace{2mm}$z$ mean & 
\vspace{2mm}Number of LAEs\\ (UDF-10, -mosaic) &  
\vspace{2mm}Expected HM12 SB$_{\mathrm{Ly\alpha}}$ & 
\vspace{2mm}Measured  SB$_{\mathrm{Ly\alpha}}$ & 
\vspace{2mm}Predicted\\ $\mathrm{\Gamma_{HI}\times f_{LLS}}$& 
\vspace{2mm}Predicted\\ f$_{\mathrm{LLS}}$ for HM12&
\vspace{2mm}Predicted\\ f$_{\mathrm{LLS}}$ for HM01&\\ [4ex]
\hline
\hline
\vspace{2mm}2.9 - 3.4 &
\vspace{2mm}3.1 & 
\vspace{2mm}33, 140 & 
\vspace{2mm}$2.0$  & 
\vspace{2mm}$<0.18$ & 
\vspace{2mm}$<0.07$& 
\vspace{2mm}$<9$\%&
\vspace{2mm}$<6$\%&\\ [2.5ex]
\hline
\vspace{2mm}3.4 - 4.5 & 
\vspace{2mm}3.9 &
\vspace{2mm}46, 242 & 
\vspace{2mm}$0.79$ & 
\vspace{2mm}$0.19\pm 0.07$ & 
\vspace{2mm}$0.11\pm 0.05$& 
\vspace{2mm}$24\%\pm 9\%$ &
\vspace{2mm}$15\%\pm 5\%$ &\\ [2.5ex]
\hline
\vspace{2mm}4.5 - 5.5 & 
\vspace{2mm}4.9 & 
\vspace{2mm}43, 162 & 
\vspace{2mm}$0.28$ & 
\vspace{2mm}$<0.14$ & 
\vspace{2mm}$<0.22$&
\vspace{2mm}$<50\%$ &
\vspace{2mm}$<35\%$ &\\ [2.5ex]
\hline
\vspace{2mm}5.5 - 6.5 & 
\vspace{2mm}5.9 & 
\vspace{2mm}16, 54 & 
\vspace{2mm}$0.1$ & 
\vspace{2mm}$<0.36$ & 
\vspace{2mm}$<1.0$&
\vspace{2mm}$<100$\% &
\vspace{2mm}$<100$\% &\\ [2.5ex]
\hline
\end{tabular}
\end{table*}
 
To estimate our constraints we consider the region between 8" to 20" and a wavelength width of $18.75\,\mathrm{\text{\AA}}$. In principle $5\,\mathrm{\text{\AA}}$ is already appropriate for an unresolved line but, given the uncertainties in the redshift estimation of the LAEs, most of them based on the \lya\ line alone, the potential emission of individual lines in a stack could be dispersed by a few hundreds of $\mathrm{km\,s^{-1}}$.

As seen in Table \ref{gobs}, the extreme assumption of a LLS \cf\ of 100\% results in upper limits of \G\ between 3 to 10 times below current constraints 
which we think to be not realistic. 
As expected, therefore, we do need to have an estimate of \flls\ in order to properly constrain the value of the UVB. This is done in Section \ref{Gammaconstraints} using the results from Section \ref{sec:simulations}.
On the other hand, by assuming a value for \G\ our observations could be used to constrain the average \cf\ of LLS
around galaxies.  

\subsection{Observational upper limits on the LLS covering fraction assuming $\mathrm{\Gamma_{HI}}$ is known}

If we assume current models of \G\ are correct, our observational results can be used to obtain an upper limit on the average \cf\ of LLSs in the selected region around our galaxies (see Table \ref{gobs}).
With this assumption, based on the HM12 UVB model we derive that LLSs cover at most 9\% and 50\% of the projected region, between 8" to 20", for $z=3.1$ and $z=4.9$, respectively, and about 24\%$\pm$9\% for $z=3.9$ in the same region. 
With the caveat that most of the obtained values are upper limits and not measurements, the decrease of \flls\ with cosmic time is consistent with the expected decrease of neutral hydrogen in the IGM due to the decreasing density (and correspondingly increasing recombination time) associated to the expansion of the Universe.

Given that currently there is a lack of observational constraints on the \cf\ at redshifts above 3 it is interesting to compare our results in the lowest redshift bin ($z=3.1$) with studies at redshifts 2 to 3. Most of them are focused on understanding how the \cf\ at CGM scales up to several virial radii around massive galaxies \citep{Rudie2012, Fumagalli2013, Prochaska2013}. Our estimated \cf\ \flls\ $<10\%$ is consistent with the low end of those studies. We think that this is reasonable since the region we have selected for our constraints goes far beyond the virial radius, our galaxies are far less massive than for most previous studies and our masking of continuum sources is very conservative.

For the annular region between 8" to 20", the predicted LLS \cf\ in the EAGLE mock cubes for the redshift bins 3.1, 3.9 and 4.9 probed with MUSE are 8.3\%, 10.6\%, and 20.2\%, respectively. For the redshift bins 3.1 and 4.9, the predicted \cf\ from the mock cubes is below the observational upper limits calculated above assuming a HM12 UVB model (9\% and 50\%). For $z=3.9$, the covering fraction inferred from the observations assuming HM12 (24\%$\pm$9\%) is more than two times above the value predicted from the EAGLE simulation. This may hint that the true value of \G\ is above the one predicted in HM12. If we use instead the model of \citet{Haardt2001}, the highest at that redshift range, we obtain that \flls\ is 15\%$\pm$5\%, consistent with the predicted value within error bars. Another explanation for the higher \cf\ may be that, even though the $l(z)$ in observations and simulations are similar (see Section \ref{coldensdist}), the slope of the radial profile of LLSs around galaxies in the simulations differs to the real one. It may be also that the selected galaxies on EAGLE represent a sample with lower column densities in their vicinities with respect to the MUSE sample. The latter is likely true for the UDF-mosaic sample because of the shallower data, which bias the selection towards more luminous and probably more massive galaxies.

\subsection{\G\ from the observations and simulation constraints on \flls}\label{Gammaconstraints}

\begin{figure}
\centering
\includegraphics[trim={1.5cm 0cm 1.5cm 0cm},clip, width=.5\textwidth]
{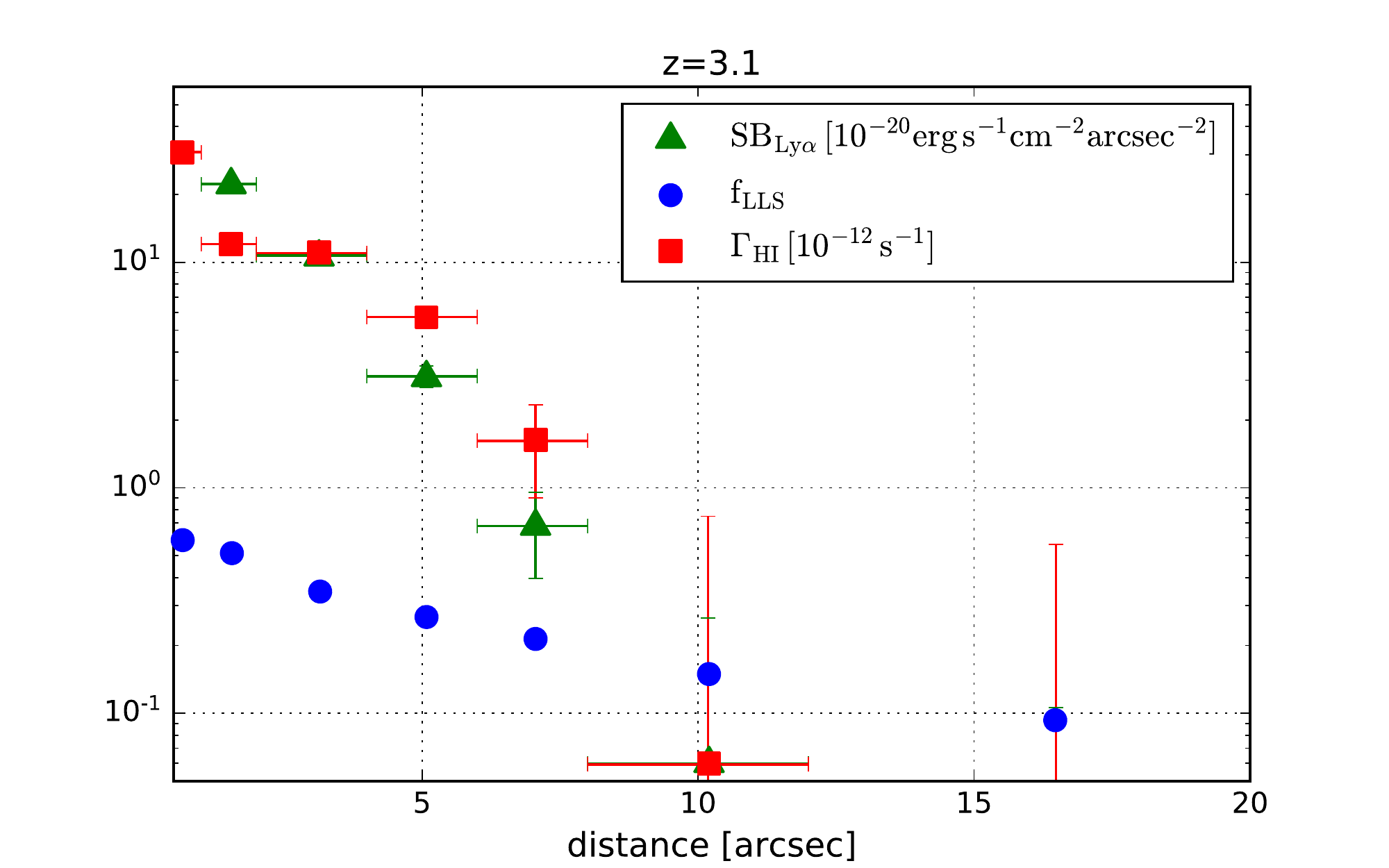}
   \caption{Radial profiles for $z=3.1$. The blue dots correspond to the expected \flls\ taken from the EAGLE mock cubes. The green dots are the measured SB at the different radial bins with a wavelength width of $18.75\,\mathrm{\AA}$. The red dots are the predicted \G\ combing the respective SB and \flls\ using Eq. \ref{gcalc}. Since there is no clear detection in this redshift bin at distances larger than 8", the \G\ values should be considered upper limits.} \label{rads1}
\end{figure} 

\begin{figure}
\centering
\includegraphics[trim={1.5cm 0cm 1.5cm 0cm},clip, width=.5\textwidth]
{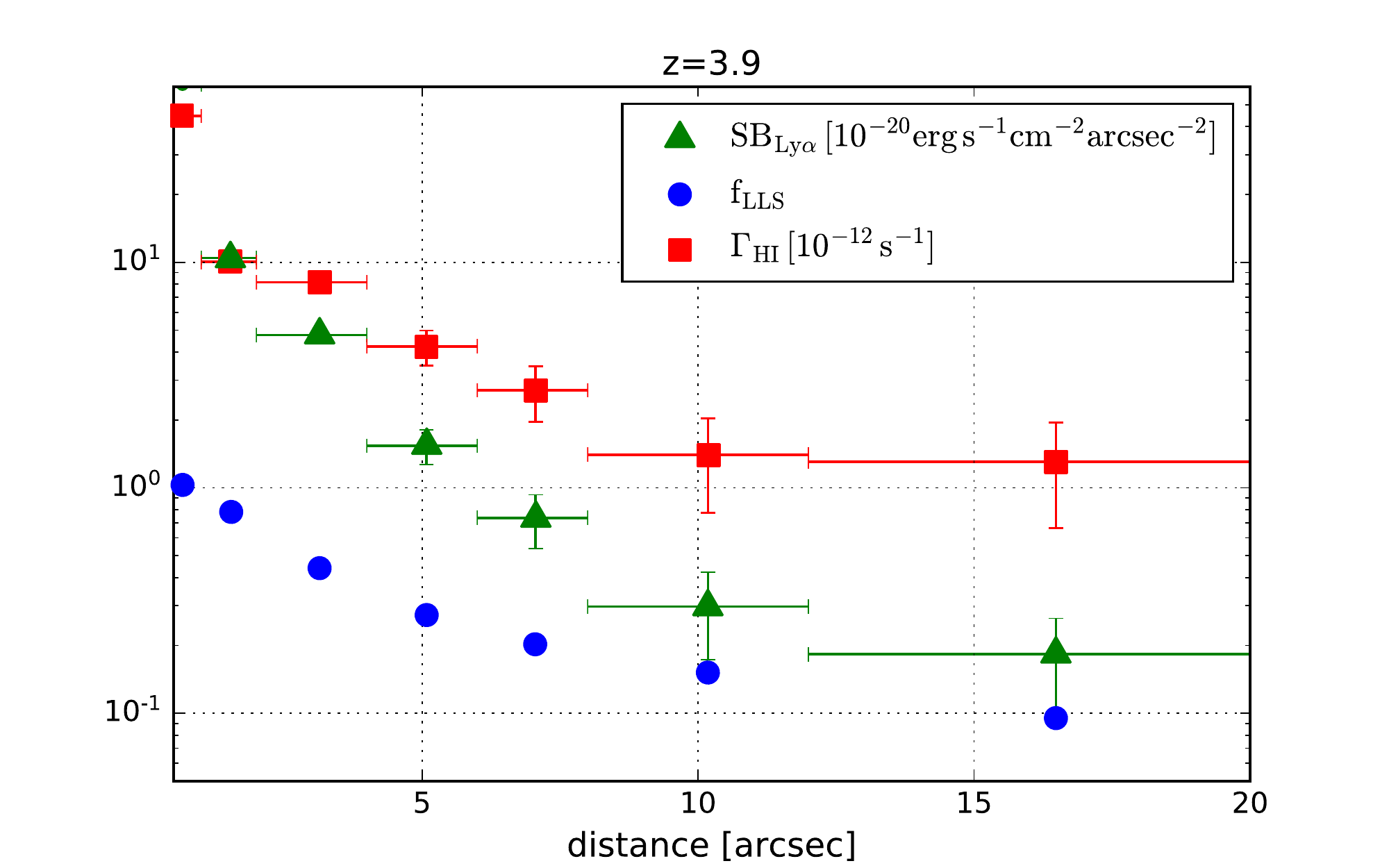}
   \caption{Same as Figure \ref{rads1} for $z=3.9$. In this case, there is a significant detection in the outer radial bins (green dots) which translates into a measurement of \G\ (red dots) using the covering fraction from EAGLE (blue dots). Although we have only two outer radial bins, the relatively constant value of the inferred \G, or, equivalently, the fact that the SB and covering fraction have the same slope at large distances could indicate that we are indeed tracing a background radiation rather than a possible galactic contribution. } \label{rads2}
\end{figure} 

\begin{figure}
\centering
\includegraphics[trim={1.5cm 0cm 1.5cm 0cm},clip, width=.5\textwidth]
{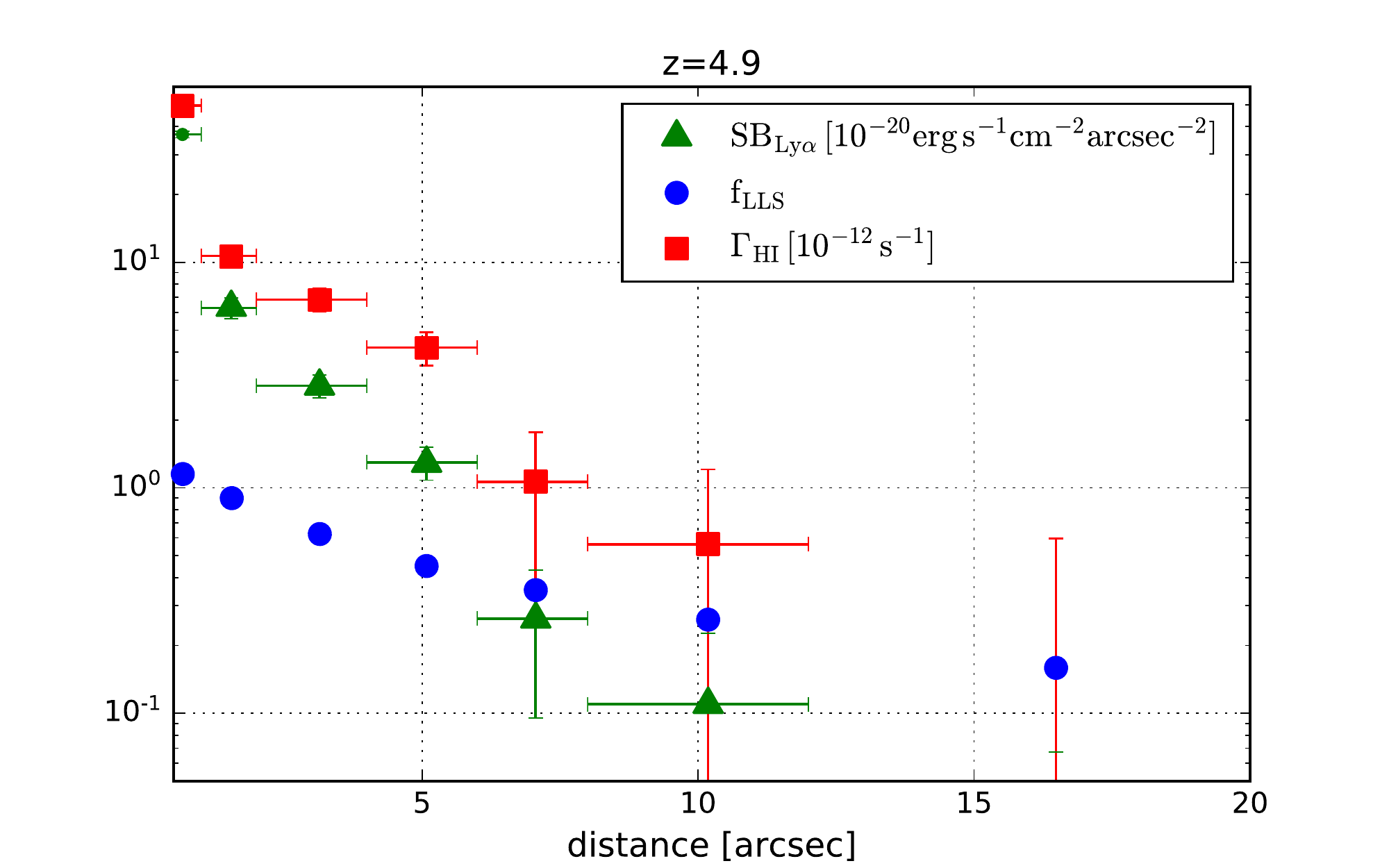}
   \caption{Same as Figure \ref{rads1} for $z=4.9$.} \label{rads3}
\end{figure} 

Using the measured SB upper limits and detection for the expected fluorescent \lya\ emission and the predicted LLS \cf\ from the EAGLE mock cubes in the same annular regions, we obtain upper limits on \G\ by combining both constraints, using Eq. \ref{gcalc}, for the 3 lowest redshift bins (Figures \ref{rads1}, \ref{rads2} and \ref{rads3}). 

Clearly, the predicted \G\ at lower radii is much higher than the average \G, produced by the UVB only, for all redshift bins, very likely due to the local ionizing radiation or \lya\ scattering produced by the central galaxy. At larger radii, where we expect the relative contribution of the UVB fluorescence to be maximal, and other mechanisms affecting the observed \lya\ negligible, we should observe that the expected \G\ flattens to a given value. That seems to be the case at $z=3.9$, where we detect \lya\ up to large radii. This result could suggest that we are indeed detecting gas clouds that are illuminated by a uniform radiation field, and therefore by the UVB.

\begin{figure}
\centering
\includegraphics[trim={.5cm 0.1cm .5cm .7cm},clip,width=.5\textwidth]{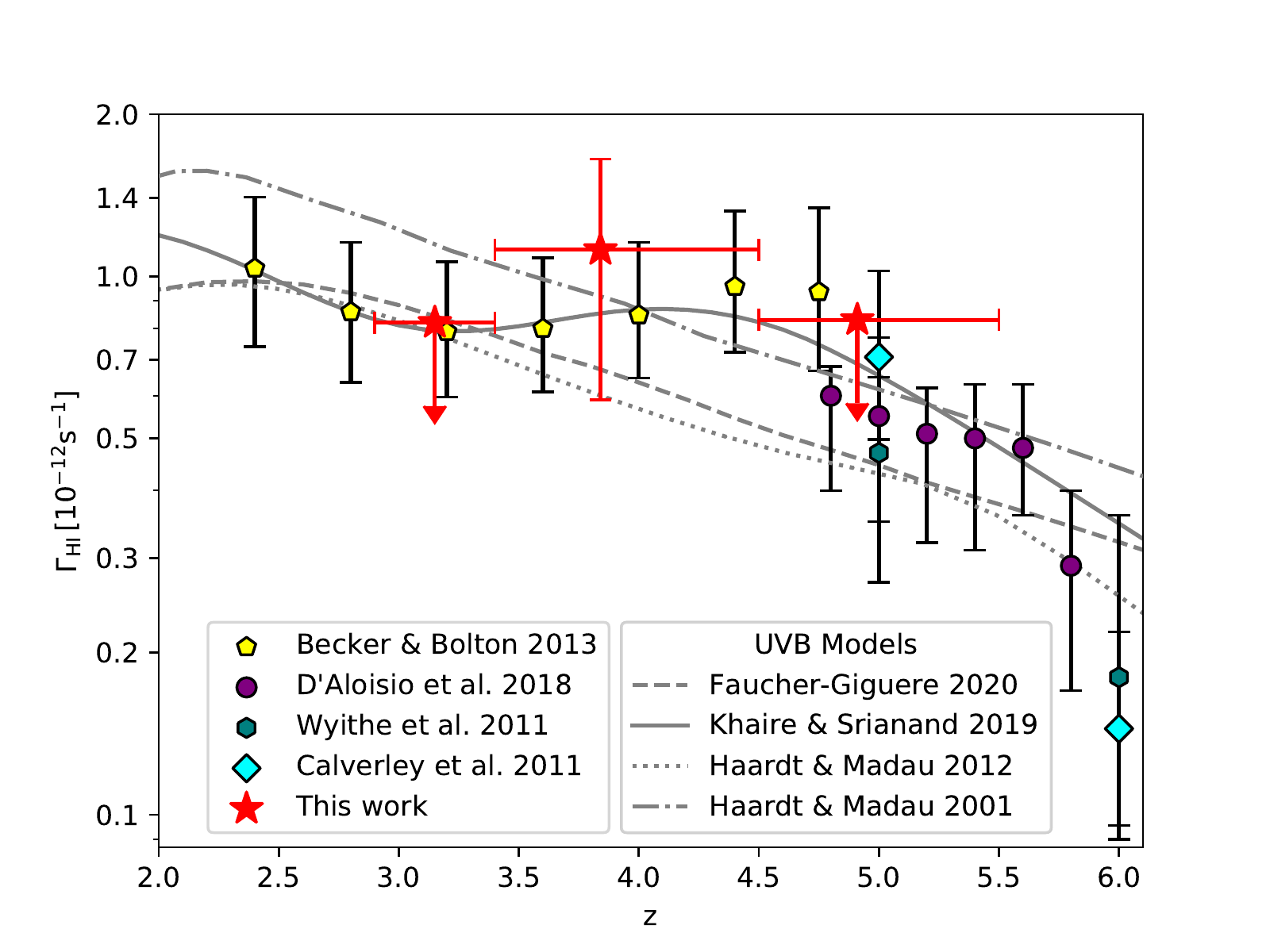}
   \caption{
   HI photoionization rates obtained with MUSE Ly$\alpha$ stacking and EAGLE simulation constraints on the covering fraction of LLS (red stars), in comparison with previous Ly$\alpha$ forest estimates (symbols) and models (lines; see legend for details). Upper limits are indicated by arrows. Model predictions and previous constraints dependent on several parameters, including the redshift evolution of the ionizing radiation escape fraction, the mean free path of ionizing photons, and the equation of state associated with Ly$\alpha$ forest clouds, while our results are mostly sensitive to the \cf\ of LLS around galaxies. Our results are in general in agreement with the Ly$\alpha$ forest constraints and strengthen the suggestion of a non-monotonic decrease of \G\ with increasing redshift which is not captured by the majority of the models (with the exception of \citet{Khaire2019} which has been designed, however, to reproduce such a trend by properly adjusting the escape fraction of ionizing photons from galaxies).}\label{uvb} 
\end{figure}

\begin{table}
\centering
\caption{Summary of the main results leading to the UVB measurements presented in Figure 9. Columns include: 1) the mean redshift for each bin; 2) the measured upper limit for \lya\ SB in units of $10^{-20}$ $\SBcgs\ $ on the annular region between 8" and 20" away from galaxies; 3) the estimated \cf\ (f$_{\mathrm{LLS}}$) over the same region as estimated from the EAGLE mock cubes; 4) the resulting \G\ upper limits and detection in units of $10^{-12}\,\mathrm{s^{-1}}$ obtained with Eq. \ref{gcalc}.}\label{uvbtable}
\begin{tabular}{| >{\centering}m{1.1cm} | >{\centering}m{2cm} | >{\centering}m{2cm} | >{\centering}>{\centering}m{2.cm}@{}  m{0cm}@{} |}
\hline $z$ mean &  Measured \\ SB$_{\mathrm{Ly\alpha}}$ & $\mathrm{f_{LLS}}$ EAGLE & Predicted\\\G\ & \\ [2ex]
\hline    
\hline
\vspace{2mm}$3.1$ & \vspace{2mm}$< 0.18$ & \vspace{2mm}$8.3\pm0.2$\% & \vspace{2mm}$<0.82$ &\\ [2.5ex]
\hline
\vspace{2mm}$3.9$ & \vspace{2mm}$0.19\pm 0.07$ & \vspace{2mm}$10.6\pm0.4$\% & \vspace{2mm}$1.12\pm 0.53$&\\ [2.5ex]
\hline
\vspace{2mm}$4.9$ & \vspace{2mm}$< 0.14$ & \vspace{2mm}$20.2\pm0.7$\% & \vspace{2mm}$<0.85$ &\\ [2.5ex]
\hline
\end{tabular}
\end{table}

As seen in Figure \ref{uvb}, our upper limits and detection at $z=3.9$ are within expectations from the previous observational constraints \citep{Calverley2011,Wyithe2011,Becker2013,DAloisio2018} and models \citep{Haardt2001,Haardt2012,Khaire2019,Faucher2020} with indications of a non-monotonic decrease of \G\ with increasing redshift, as was also suggested by Becker \& Bolton (2013).  
A summary of our UVB constraints is presented in Table \ref{uvbtable}.


\section{Discussion}\label{sec:discussion}

As we have shown in the previous section, our method is able to provide direct observational constraints on the product of the UVB photoionization rate and the covering fraction of LLS at a given distance from galaxies. By assuming the covering fraction of LLS as given by the EAGLE cosmological simulation we have been able to put some constraints on the value of \G\ that are independent from previous measurements and which are consistent with the majority of them.
In particular, we have obtained two upper limits and a possible measurement in our median redshift bin. 
Within the assumptions and caveats of our method, discussed in detail below, this seems to suggest that the intensity of the UVB could be not monotonically decreasing within redshift in the range 
$3<z<5.5$, contrary to what suggested by the majority of the theoretical estimates of the UVB. 
Interestingly, our result seems compatible with the independent estimate made by Becker \& Bolton (2013)
and later reproduced in the model of \citet{Khaire2019} (which has adjusted the escape fraction in order to
match the data). Our upper limit at $z\approx3$ also
seems to strengthen the suggestion made by previous studies that the \citet{Haardt2001}
model produces a \G\ that is too high at this redshift. 

In the following, we discuss the main uncertainties and limitation in our method and analysis. 

\subsection{Physical mechanisms affecting the observed \lya}

One of the basic assumptions in our calculation of \G\ is that the only mechanism to produce \lya\ photons at distances above 8" from galaxies is fluorescence from the UVB. Other mechanisms potentially affecting the expected \lya\ emission include scattering, collisional excitation, local sources of ionizing or \lya\ photons, and absorption by dust. The temperature, density and ionization state of the medium, and the properties of local sources of \lya\ and ionizing photons, will mostly determine the relative importance of each process and therefore the expected rate of \lya\ photons observed. Most of these effects increase the expected SB, and therefore we conclude that the real value of \G\ is very likely not higher than our estimated upper limits and detection. Our main conclusions are therefore mostly unaffected by these uncertainties.

\subsubsection{\lya\ scattering}
 
The expected \lya\ SB could be increased by \lya\ photons escaping from HII regions within the galaxies, and subsequently scattering across the CGM/IGM until being directed toward the observer. 
For \lya\ scattering to be efficient up to very large distances it would be necessary an optically thin to moderately optically thick medium to \lya\ photons over several tens of kpc (i.e. a medium with an optical depth $\tau\sim1$). Indeed, within very optically thick media, Ly$\alpha$ photons mostly diffuse in frequency rather than in space. Unless the velocity field around galaxies is properly arranged to produce such a relatively constant optical depth, we think that such a situation is difficult to be achieved for a large fraction of galaxies and be relevant therefore for our stacking analysis. In any case, even assuming that scattering does contribute to the emission our inferred upper limits on \G\ and their interpretation would remain valid. If our detection at $z\approx3.9$ is due to the contribution of scattering instead of UVB fluorescence, this would require a (unknown) mechanism that is able to boost scattering preferentially at this redshift.  

\subsubsection{Collisional excitation}

Unlike recombination, collisional excitation is extremely sensitive to the temperature and requires a partially ionized medium to occur. In the CGM/IGM, assuming the neutral gas is confined in self-shielded clouds surrounded by an ionized medium, 
we expect some contribution of collisional excitation coming from the transition between the ionized exterior and the neutral interior of the clouds. The modelling of this process is very sensitive to several physical parameters and large variations could be expected depending on temperature and density distribution. Although redshift could also play some role, we are not aware of any mechanism that could be able to boost this radiation at $z\sim4$ with respect to the other explored redshifts. Also in this case, if collisional excitation do contribute, this does not change the interpretation of our upper limits. 
    
\subsubsection{Ionizing radiation from the central galaxies} 

Similarly to \lya\ scattering, ionizing photons escaping from our selected galaxies could contribute to the total photoionization rate, and we expect their contribution to decay at least by the inverse-square law, therefore being less relevant the farther away we are from galaxies. 
How much do they contribute it is however difficult to estimate without a direct constraint on both their ionizing $f_{esc}$
and their production rate of ionizing photons. UVB models do assume that galaxies are the dominant sources of ionizing photons
in the universe above $z>3$ and (arbitrarily) model their $f_{esc}$ as monotonically increasing with increasing redshift 
between $3<z<6$ (this is necessary in these models to compensate the sharp decline in AGN number densities at high redshift and to fit reionization constraints). In this scenario, it would be difficult to explain our non-monotonic behaviour of \G\ with redshift and our detection for the median redshift bin as due to an increased contribution from the central galaxies. Our conclusion concerning the upper limits would be of course unchanged even considering a possible contribution from local sources. 

\subsubsection{Contribution from AGN}\label{AGNcontr}

We expect that AGN activity could be especially important to provide additional ionizing photons, possibly on large scales depending on the AGN ionizing luminosity. Could our detection in the redshift bin $3.4<z<4.5$ be affected by an enhanced \G\ due to contribution of some AGN (either in the central galaxies or in their proximity)? 

Within this redshift range, several AGN have been detected within the UDF-mosaic area or in contiguous regions, particularly through their X-ray emission \citep[e.g.][]{Luo2017}. The majority of the AGN with spectroscopic confirmation cluster in the redshift range $3.6<z<3.8$, which we have excluded and treated separately as discussed in Sec. \ref{AGNeffect}. Apart from issues related with the spectroscopic confirmation of the detected AGN,
current catalogues suffer from severe incompleteness at these redshifts already at moderate intrinsic luminosities due to obscuration along our line of sight \citep[e.g.][]{Vito2018}. Such obscuration does not exclude however that ionizing radiation is emitted along other directions, as seen in several cases for type-II AGN \citep[e.g.][]{Brok2020}. Moreover, AGN are highly variable sources in all bands, including X-ray, on scales that can be as short as a few days, while the persistence of their observable effect on fluorescent emission depends on time scales that are as large as the recombination time of the illuminated gas, i.e. several Myr. For all these reasons, a simple cross-check of the \citet{Luo2017} catalogue is not sufficient to exclude the possible contribution of AGN to the \G\ in a specific redshift range and area and we resort instead to the following statistical approach. 

We first derive the expected number density of AGN per unit area in our redshift range ($\Sigma_{\mathrm{AGN}}$), corrected for incompleteness, using the results of \citet{Vito2018}, which are based on the Chandra Deep Field South (containing the UDF-mosaic area). In particular, they have obtained that the volume density of AGN at $z\sim3.9$ above a luminosity of log$(L_X/[\mathrm{erg/s}])>42.5$ in the 2-10 keV band, corrected for incompleteness, should be around $10^{-4}$ cMpc$^{-3}$ 
(this number decreases by a factor of about two if the incompleteness correction is not included\footnote{This incompleteness correction is mostly relevant for obscured AGN along our line of sight which in \citet{Vito2018} are defined as sources with $\mathrm{log(N_H\,[cm^{-2}])}>23$}). 
The comoving volume per square arcmin in the redshift range $3.4<z<4.5$ is around $3.5\times10^3$ cMpc$^3$ (using the conversion factor between area on the sky and physical distance at z=4). Therefore, we do expect a $\Sigma_{\mathrm{AGN}}\sim0.35$ arcmin$^{-2}$ in our redshift range. 

Next, we estimate what is the area of the proximity region produced by these AGN, which depends on the radius at which the AGN ionizing flux is at least equal to the UVB one. This area would of course scale linearly with the ionizing luminosity of the AGN and there would be therefore a distribution of values for the proximity regions. For our statistical analysis however, it would be sufficient to obtain the average area. Thus, for simplicity, we assume that all AGN have the average luminosity in the 2-10 KeV band as in the X-ray sample of \citet{Vito2018}. Using their X-ray Luminosity Function within the range $(42.5<\mathrm{log}(L_X/[\mathrm{erg/s}])<45)$ (at higher luminosities the Luminosity Function drops steeply), we obtain an average luminosity in the 2-10 keV band of about log$(L_X/[\mathrm{erg/s}])\sim43.7$. This luminosity is corrected for intrinsic absorption. 
We transform this intrinsic X-ray luminosity into a luminosity at the Lyman Limit ($L_{\mathrm{LL}}$) using the linear 2-10 keV X-ray to bolometric correction and the bolometric to 1450$\mathrm{\AA}$ correction factors of \citet{Runnoe2012}, and, finally, the 1450$\mathrm{\AA}$ to 912$\mathrm{\AA}$ correction factor of \citet{Lusso2012}. The resulting average luminosity at the Lyman Limit is $L_{\mathrm{LL}}\sim2\times10^{29}$ erg s$^{-1}$ Hz$^{-1}$. All these corrections are associated with a large error (and still unclear dependencies on redshift, AGN type, luminosity, and spectral shape) which we neglect here for simplicity, therefore this conversion should be considered as an order of magnitude estimate. Moreover, we have assumed that the ionizing luminosity is emitted isotropically, while it is highly probable, especially for obscured AGN along our line of sight, that the emission is oriented along the plane of the sky \citep[see e.g.,][for a few examples, one of which in the UDF-mosaic area]{Brok2020}. In the case of anisotropic emission, since more radiation is emitted in some particular direction, the proximity regions would be larger, e.g., along the ionization cones, but absent in other directions. These two effects could somewhat balance each other in terms of area and we will neglect them here for simplicity. 

The \G\ due to an AGN with a power-law spectrum of the form $f_{\nu}\propto f_0 \nu^{\alpha}$, in function of distance from the source (assumed to emit isotropically) is:

\begin{equation}
\mathrm{\Gamma_{HI}(AGN)}\simeq 2\times10^{-12} L_{\mathrm{LL,30}}\cdot R^{-2}\cdot \frac{4.45}{2.75-\alpha},
\end{equation}

where $L_{\mathrm{LL,30}}$ is the luminosity at the Lyman Limit in units of $10^{30}$ erg s$^{-1}$ Hz$^{-1}$, $R$ is the distance in units of physical Mpc. Using the estimated luminosity above, a value of $\alpha=-1.7$ \citep{Lusso2012} and considering the \G\ from the HM12 model (which corresponds to about $0.6\times10^{-12}$ at $z=4$) as a reference value, we obtain a radius of the proximity region of about 0.8 physical Mpc (pMpc) or about 4 comoving Mpc (cMpc) at $z\simeq3.9$. This distance corresponds to a circular, projected area of about 11 arcmin$^{2}$ for isotropic emission, which is a relatively large area, much larger than a single MUSE FoV and larger than the UDF-mosaic area. As a reference, in the assumptions above, the radius of the proximity region is simply connected to the absorption corrected 2-10 KeV X-ray luminosity ($L_X$) by the following relation

\begin{equation}
R_{\mathrm{eq}}^{\mathrm{HM12}}\simeq 0.8 \times \left(\frac{L_X}{5\times10^{43}\ \mathrm{erg}\ \mathrm{s}^{-1}}\right)^{1/2} \mathrm{pMpc}
\end{equation}

Combining the average area of the AGN proximity region with the value of $\Sigma_{\mathrm{AGN}}$ implies that each line of sight in the redshift range $3.4<z<4.5$ is expected to cross, on average, $N\simeq4$ AGN proximity regions, i.e. regions in which the \G\ is dominated by a local AGN rather than the UVB. This is also the same number that is expected if the observed FoV is smaller than the proximity region area, which is similar to our case. We stress that there is a large error associated with this number due to the several simplifications made but also due to cosmic variance and possible clustering effects that are especially relevant for small field of view observations.

 It is difficult to estimate what would be the boost factor in \G\ value associated with these proximity regions, as it would depend on several geometrical factors that are difficult to assess with a low number statistic. However, it is interesting to note that our simple calculation showed that across such a large redshift range there is a non negligible probability of finding a few regions in which the \G\ is boosted with respect to the average UVB value and such regions are not necessarily associated with a detectable AGN (either because of obscuration effects along our line of sight or short-term variability but also because the AGN could be outside of the FoV).
 
 Taking the number calculated above at face value and considering the $\Delta$z associated with a proximity region would imply that within about 40 MUSE layers in our redshift range $3.4<z<4.5$ there could be a local enhancement in the \G\ due to AGN. Assuming a uniform distribution of galaxies, this would imply that about 4\% of our galaxies are affected. Even considering possible clustering of galaxies in proximity of the AGN, this number is still small enough to not bias significantly our stacking analysis. However, our calculation hints at the necessity to build a large sample of galaxies and to follow a blind statistical approach, i.e. not based on a few individual detections even far away from galaxies, in order to estimate the average \G\ due to the UVB alone. 

\subsubsection{Ionizing radiation and \lya\ emission from undetected galaxies}

Considering the uncertainties in the slopes of the UV and \lya\ luminosity functions at low luminosities, undetected galaxies surrounding our selected galaxies could potentially contribute to both the \lya\ and ionizing photons. Some theoretical models and observational results do suggest that less luminous galaxies could have a higher UV $f_{\mathrm{esc}}$. Because we perform a stack and consider a
large volume around galaxies in average regions of the universe, if undetected galaxies really contribute to the local \G\, then we can consider their ionizing photons as a part of the total UVB budget. The situation is different however if they contribute to the \lya\ emission and not to the \G. As discussed in the previous cases, however, this would not affect our conclusions which are mostly based on upper limits and would remain valid even including this contribution. 

\subsubsection{\lya\ absorption by dust}
    
Among the few mechanism that could bias in the opposite direction our estimate of \G\ there is the possibility that Ly$\alpha$ photons resulting from UVB fluorescence are absorbed by dust and therefore missing in our measurement. However, if the majority of LLSs that are emitting fluorescent radiation are Intergalactic, as assumed here (consistently with current numerical simulations), we expect very little to no dust present in these systems.

\subsection{$\mathrm{\Gamma_{HI}}$ around galaxy and AGN overdensities: the effect of local sources}\label{AGNeffect}

\begin{figure}
\centering
\includegraphics[trim={0.3cm 0cm 0cm 0cm},clip, width=.45\textwidth]
{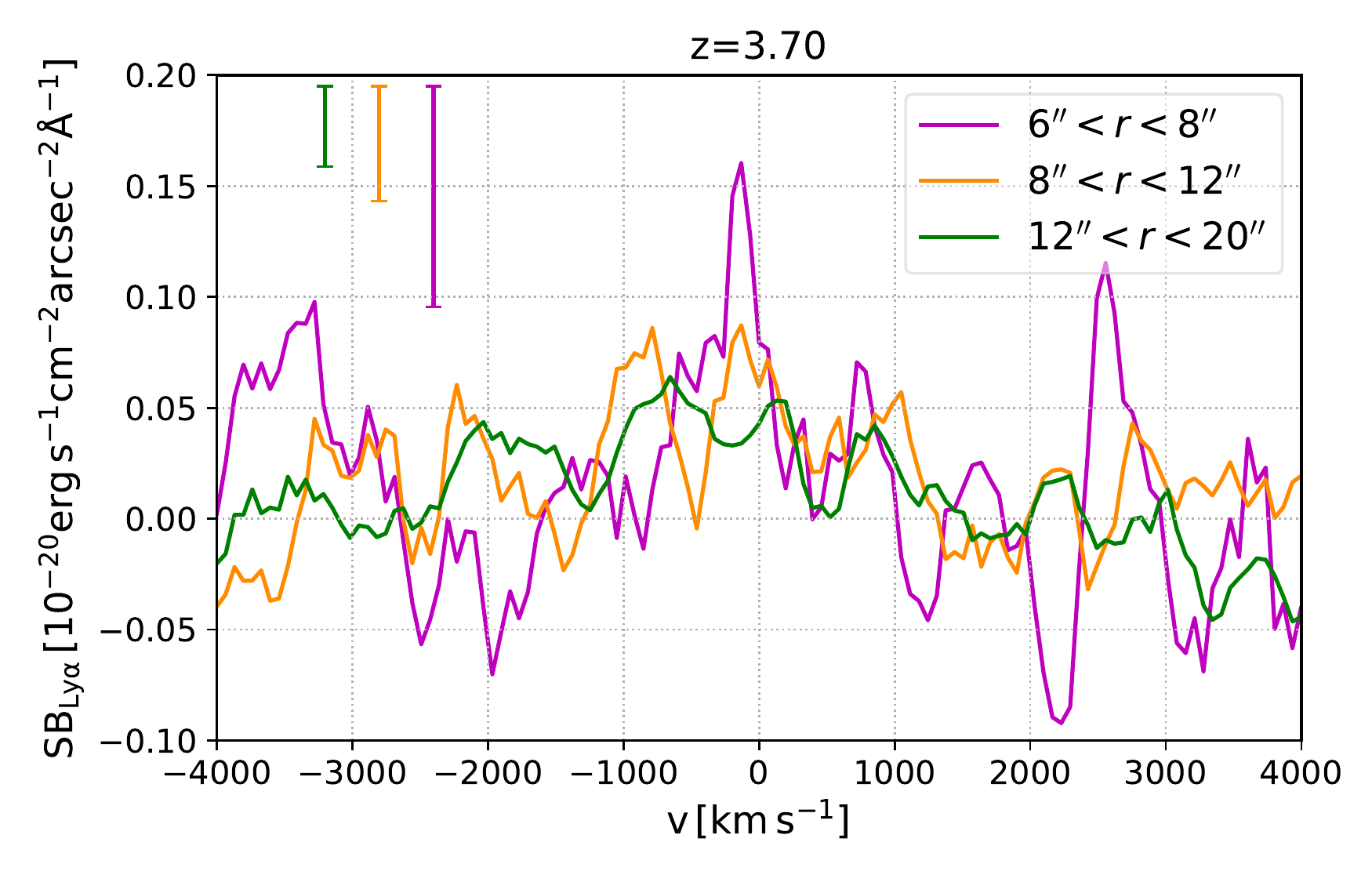}
   \caption{Stacked spectra extracted from three radial annuli around LAEs in the overdensity found at $z=3.7$. }    \label{ovspec}
\end{figure}

\begin{figure}
\centering
\includegraphics[trim={1.5cm 0cm 1.5cm 0cm},clip, width=.5\textwidth]
{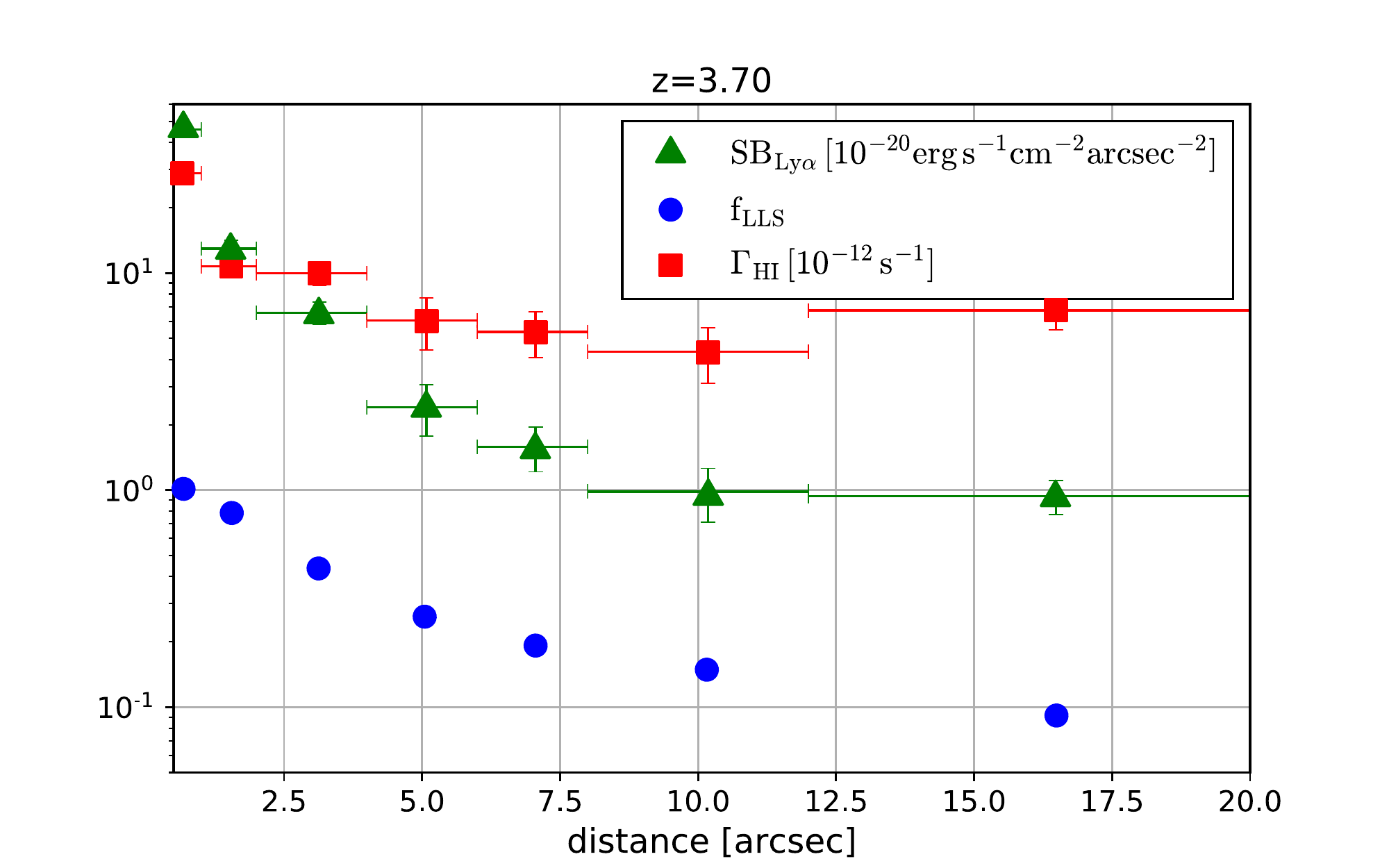}
   \caption{Similar to Figure \ref{rads1}, but for the overdensity found at $z=3.7$.}    \label{ovprof}
\end{figure} 

As discussed above, we do expect that some regions should have an enhanced \G\ due to the presence of AGN and increased
clustering of galaxies. One of these regions is the detected overdensity in the UDF-mosaic at $z=3.7$, which we have masked in our analysis to avoid any possible bias in our measurement. 
Within the redshift range $3.65<z<3.75$ there are 18 LAEs for UDF-10 and 45 for UDF-mosaic. These numbers corresponds to a modest overdensity of only a factor 2.3 with respect to the full redshift range considered in our median bin. 
 However, there are four spectroscopically confirmed and relatively bright AGN close or within the $3.65<z<3.75$ range and within about 7 arcmin from the center of the UDF-10 and the UDF-mosaic \citep{Luo2017} which do constitute a significant overdensity. 
 We note that the two brightest AGN, which are classified as type-II and which also show extended and bright Ly$\alpha$ nebulae \citep{Brok2020}, are outside of the UDF-mosaic area \citep[but within the MUSE-Wide survey area,][]{Urrutia2019} and therefore their nebulae do not enter in our stacking analysis.  
 Their intrinsic-absorption corrected X-ray luminosities (0.5 to 7 keV) are between $10^{44}$ and $5\times10^{44}$ erg s$^{-1}$,  i.e. up to ten times brighter than the average X-ray luminosity considered in Sec. \ref{AGNcontr} (although the considered band is slightly different). Therefore, they could be associated with (possibly overlapping) proximity regions extending up to about a few tens of arcmin from the brightest source, clearly including the UDF-mosaic area.
The spectra for the outermost radial bins and the radial profiles can be seen in Figures \ref{ovspec} and \ref{ovprof}. 

There is a clear detection of \lya\ emission at all radial bins. From the outermost bins and using the \cf\ estimated from EAGLE 
we derive a $\Gamma_{\mathrm{HI}}\simeq6\times10^{-12}$, which is about 5 times higher than the measured \G\ in the redshift bin $3.5<z<4.5$ and 10 times higher than expected using the Haardt \& Madau (2012) model. This large boost is fully consistent with our expectations as discussed in Sec. \ref{AGNcontr}. The detection of this signal has been likely possible both for the enhanced \G\ over large scales and for the presence of a large number of galaxies, and therefore stacking elements, within such large AGN proximity regions.   

 We notice that the \G\ estimate above is based on the \cf\ derived from the simulations which do not consider the effect of local ionizing sources such as AGN (discussed in Sec. \ref{sim:localUV}) but also the possible increase in \cf\ due to clustering effects. 
  We could try to estimate the latter effect, for instance, taking as a reference the correlation found for the distance to the fifth neighbour ($\mathrm{d_{5th}}$) in Fig. \ref{n5d}. In this case, the \cf\ could be increased by as much as 20\% with respect to the mean. Clustering effects could therefore imply that the predicted \G\ could be lower by 20\%, which is still consistent with an enhancement 
  due to local sources and in particular due to the presence of the detected AGN.

Finally, it is important to notice that our method is only able to give constraints on the integrated emission of the UVB 
which should be assumed in order to translate this information into a value of \G.
The presence of local ionizing sources and especially potential AGN emission in some particular regions, could significantly change the shape of the UV spectra and therefore the inferred \G. 

\subsection{Prospects and strategies for the future}

\begin{figure}
\begin{center}
\includegraphics[trim={.3cm .1cm .1cm .2cm},clip,width=3.5in]{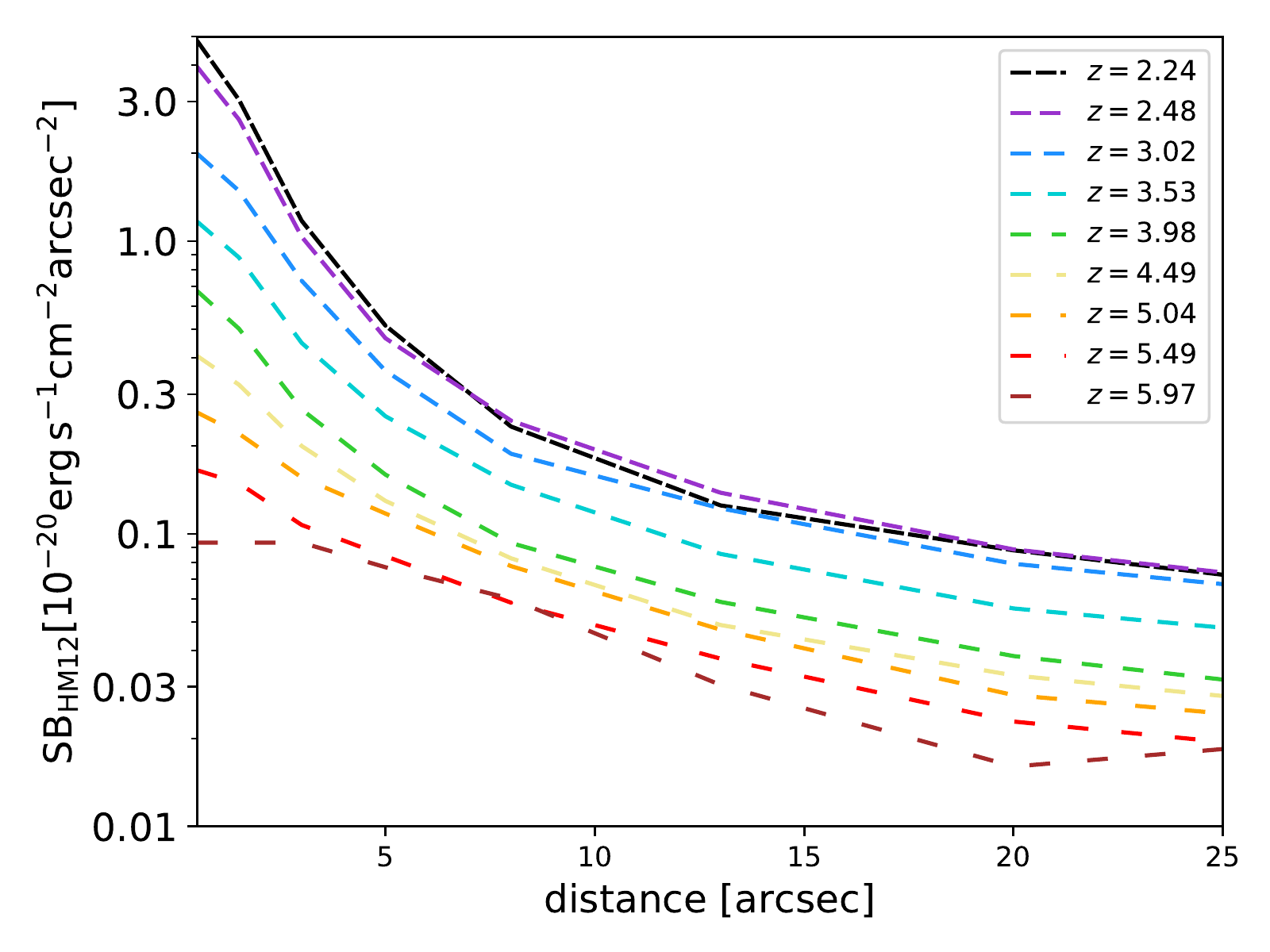}
   \caption{Predicted \lya\ SB profiles at different redshifts. SB values were obtained by assuming a HM12 UVB photoionization model combined with EAGLE predictions for the LLS covering fraction.} \label{SB12}     
\end{center}
\end{figure} 

New deep observations can improve our constraints on \G\ using \lya\ emission by increasing the sample of galaxies, exposure time, and extending our studies to lower redshifts. In this context, KCWI, and HETDEX IFU observations can extend our studies down to $z\sim2$, by benefiting from the decreasing effect of the SB dimming compared to MUSE \lya\ observations, and the connection with current H$\alpha$ observations which could disentangle the physical mechanisms responsible for the observed \lya\ emission. In Fig. \ref{SB12}, we predict how the \lya\ SB radial profiles around galaxies should look like assuming the HM12 model for the UVB and \flls\ profiles derived from EAGLE, which suggest the optimal detection redshift is between $2.2<z<2.5$. For KCWI \citep{Morrissey2018}, we benefit from a $2\sigma$ sensitivity of $7\times10^{-20}\,\SBcgs\ $ in about 4 hours at $4000\,\mathrm{\text{\AA}}$, although its small F.o.V. would probably require targeting specific regions, especially around galaxies in more clustered environments as suggested by our analysis in Section \ref{sec:galprops} and Appendix A. Large statistical sample would be needed in any case to overcome the effect of local \G\ enhancements, as discussed in the previous section, due to local sources and especially AGN, which become also more abundant at lower redshift.

\section{Summary}\label{sec:summary}

The cosmic UV background shapes the ionization state and thermal evolution of the Intergalactic Medium (IGM), which is thus a crucial ingredient of cosmic structure formation models. Most previous studies attempting to measure the UVB focused primarily on characterizing the properties of the \lya\ forest by comparison with numerical simulations or through theoretical models which assume parameters that are very difficult to estimate, such as the redshift evolution of the escape fraction of ionizing photons from faint galaxies.

Here we presented the results from an independent method to constrain the UVB and in particular, the photoionization rate of hydrogen $\mathrm{\Gamma_{HI}}$, based on the detection of fluorescent \lya\ emission produced by Lyman Limit Systems (LLS) illuminated  by the UVB at $3<z<6.5$. In order to achieve the required sensitivity limits, we performed a three-dimensional stacking analysis of the IGM around about 700 \lya\ emitters detected within some of the deepest MUSE fields available to date.  
In particular, our method provides direct observational constraints on the product of the ionizing radiation field intensity (\G) and the covering factor of LLSs around our galaxies. Before performing our stacking analysis, we have divided our galaxy sample into four redshift bins roughly centered at $z\approx3.1$, $z\approx3.9$, $z\approx4.9$ and $z\approx5.9$ and we have excluded regions associated with clear galaxy and AGN overdensities for which the inferred \G\ is likely affected by local sources, as we have discussed and demonstrated in Sec. \ref{AGNcontr} and \ref{AGNeffect}.

Our results can be summarized as follows:

\begin{itemize}
    \item \lya\ emission is typically detected up to 8" away from galaxies in our stacks at all redshifts. We consider this emission as likely arising from processes related to the galaxies themselves and their CGM rather than fluorescence from the UVB, consistently with previous studies. 
    \item \lya\ emission within 8" and 20" away from galaxies is detected at $z\approx3.9$ at a SB level of $0.19\pm 0.07$ times $10^{-20}$$\SBcgs\ $, while for the other redshift bins we were only able to obtain upper limits to the SB (0.18, 0.16 and 0.36 times $10^{-20}$$\SBcgs\ $ at  $2\sigma$ for redshifts 3.1 and 4.9 and 5.9, respectively). We have not considered the $z\sim5.9$ further given the poor constraint due to the low number of galaxies in the stacking analysis. 
    \item Assuming that the emission between 8" and 20" is mainly powered by UVB fluorescence, these SB translated into a value of $\mathrm{\Gamma_{HI}\times f_{LLS}}\approx0.19\pm0.06\times 10^{-12}\mathrm{s^{-1}}$ at $z=3.9$ and the following upper limits: $0.07\times 10^{-12}\mathrm{s^{-1}}$ for $z=3.1$,  $0.11 \times 10^{-12}\mathrm{s^{-1}}$ for $z=4.9$. These are direct observational constraints from which a value of $\mathrm{\Gamma_{HI}}$ could be derived from future observational measurements of the covering fraction of LLS (\flls) around galaxies at these redshifts. 
    \item In absence of observational constraints on \flls\ at these redshifts, we have combined our results with the LLSs \cf\ estimated from the EAGLE cosmological simulation, obtaining the following values or ($2\sigma$) upper limits on $\mathrm{\Gamma_{HI}}$: $<0.82$, 1.12$\pm$0.53 and $<0.85$ times $10^{-12}\mathrm{s^{-1}}$ for redshifts 3.1, 3.9 and 4.9, respectively. These values are consistent with the majority of other recent independent constraints (and in tension with the HM01 model at $z\approx3$) and suggest a non-monotonic decrease of $\mathrm{\Gamma_{HI}}$ with increasing redshift between $3<z<5$ as also suggested by some other indirect measurements and models.
    \item Assuming instead a value of \G, we used our detected \lya\ emission values and upper limits to obtain the \cf\ of LLSs around LAEs. In particular, assuming the HM12 UVB model, we derive a \cf\ of $<9\%$, 22\%$\pm$8 and $<41\%$ for redshifts 3.1, 3.9 and 4.9, respectively, within 150 kpc from our LAEs. We notice that our results at $z=3.1$ are consistent with previous studies at redshifts $2<z<3$ \citep[e.g.][]{Rudie2012}.
    \end{itemize}
    
Despite the current observational and theoretical limitations, which we have discussed in this work, we have shown that the available and upcoming observational datasets are in principle deep enough to provide new constraints on the value of the UVB background at high redshift. Moreover, we have shown that local sources of radiation and especially AGN are expected 
to enhance the local radiation field in a non-negligible fraction of the volume probed by MUSE datacubes, 
hinting at the necessity of using a statistical approach and a large sample of stacking elements in order to avoid a possible bias.
At the same time, new observational constraints on the covering fraction of LLSs at $z>3$ in function of distance from galaxies are vital to reduce the possible uncertainty on the UVB estimated with the method and assumptions presented here. 
When combined to the upcoming ultra-deep datasets or large surveys expected to be completed soon, these observations could provide the strongest available constraints on the UVB and thus on the production and escape of ionizing photons from the whole population of star forming galaxies and AGN at high redshift.  

\section*{Acknowledgments}
 
This research made use of Astropy, a community developed core Python package for Astronomy (Astropy Collaboration et al. 2013). topcat, a graphical tool for manipulating tabular data, was also utilized in this analysis (Taylor 2005). SG would like to thank Nastasha Wijers for the discussion on the column density distribution in EAGLE. SC gratefully acknowledges support from Swiss National Science Foundation grants PP00P2\_163824 and PP00P2\_190092, and from the European Research Council (ERC) under the European Union's Horizon 2020 research and innovation programme grant agreement No 864361. GP acknowledges support from the Swiss National Science Foundation
(SNF) and from the Netherlands Research School for Astronomy (NOVA).

\section*{Data Availability}
The data underlying this article will be shared on reasonable request to the corresponding author.

\bibliographystyle{mnras}
\bibliography{refs}

\appendix
\counterwithin{figure}{section}

\section{EAGLE radial profiles for different galaxy properties}

\begin{figure*}
\centering\includegraphics[width=.49\textwidth]
{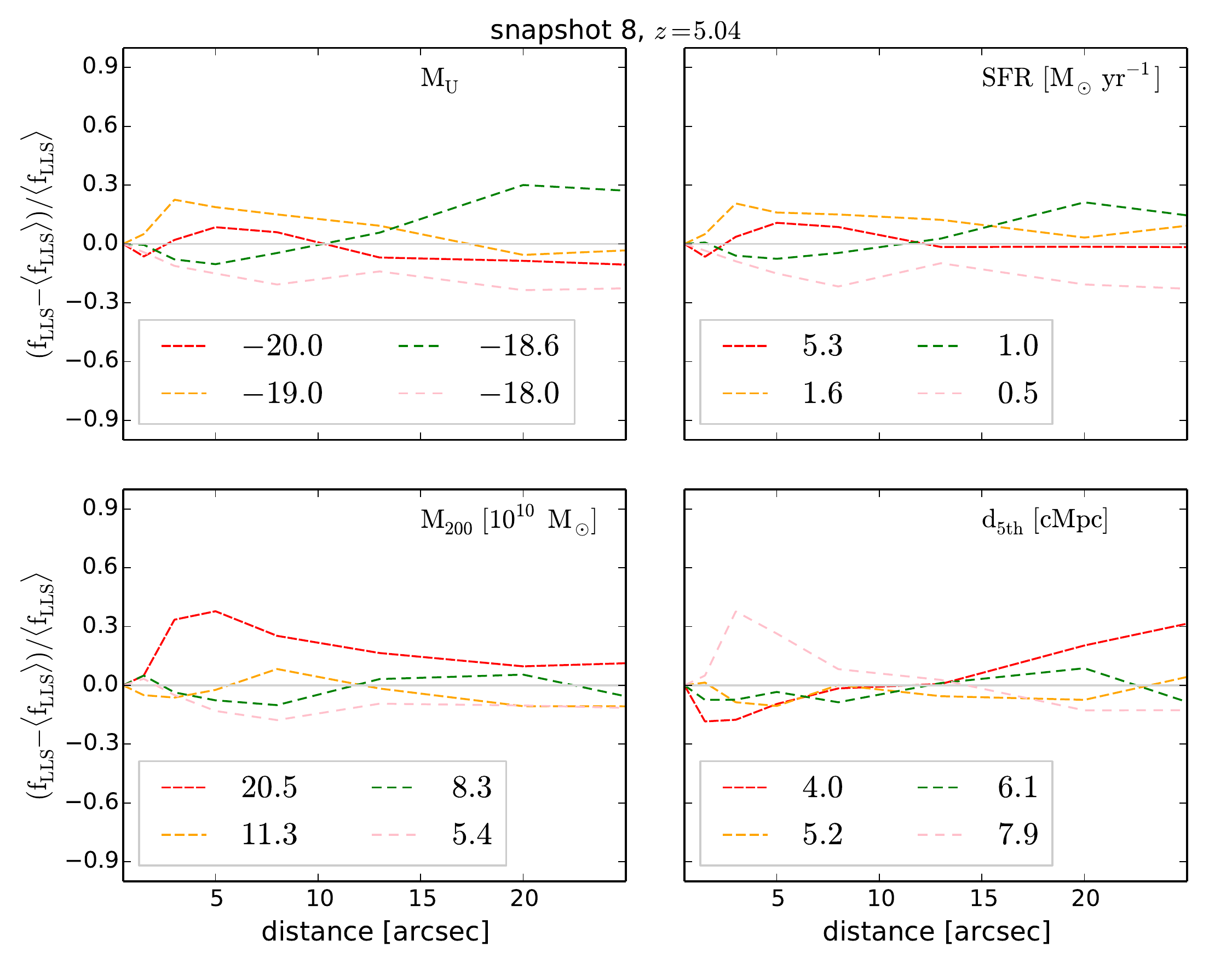}
\includegraphics[width=.49\textwidth]
{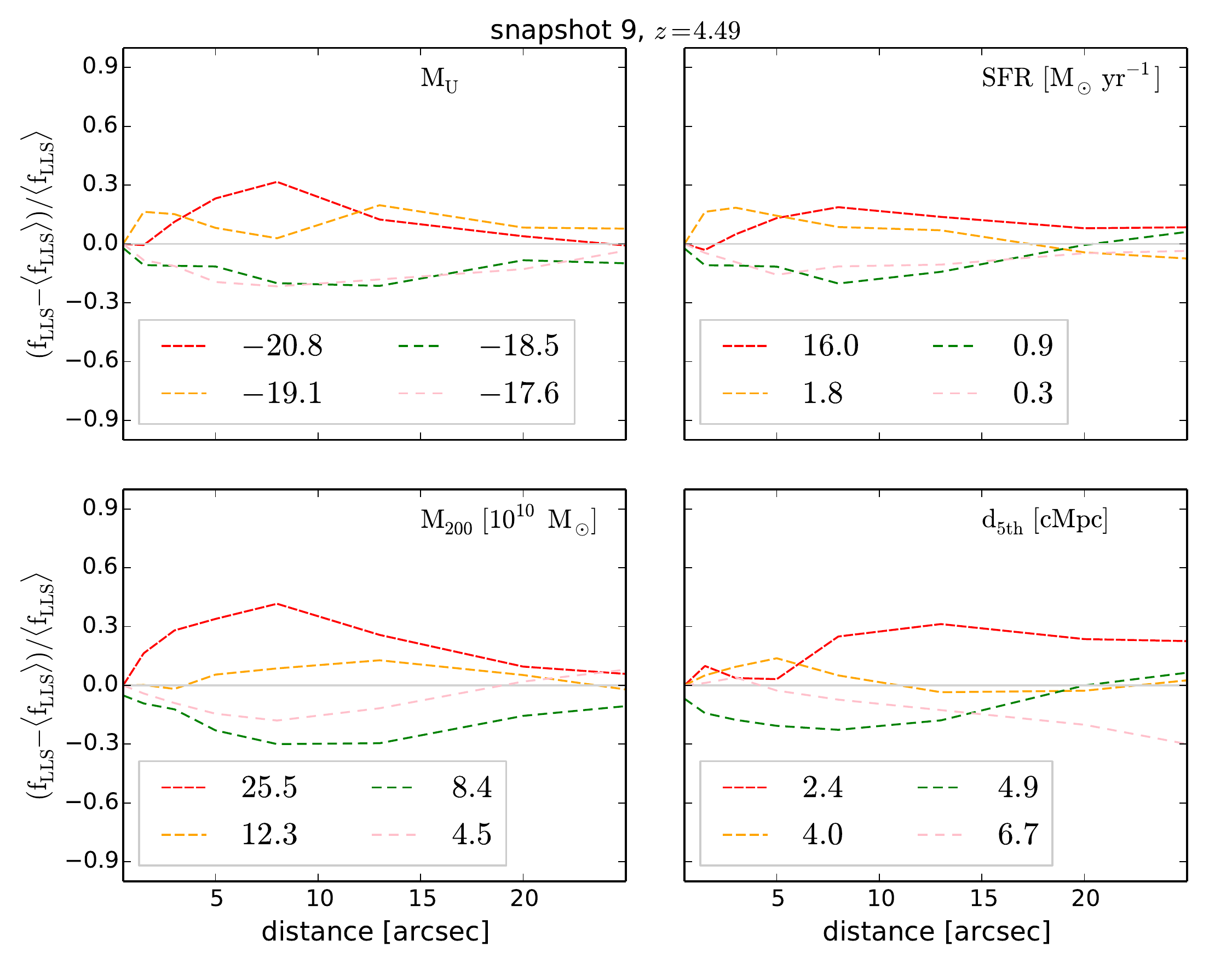}
\includegraphics[width=.49\textwidth]
{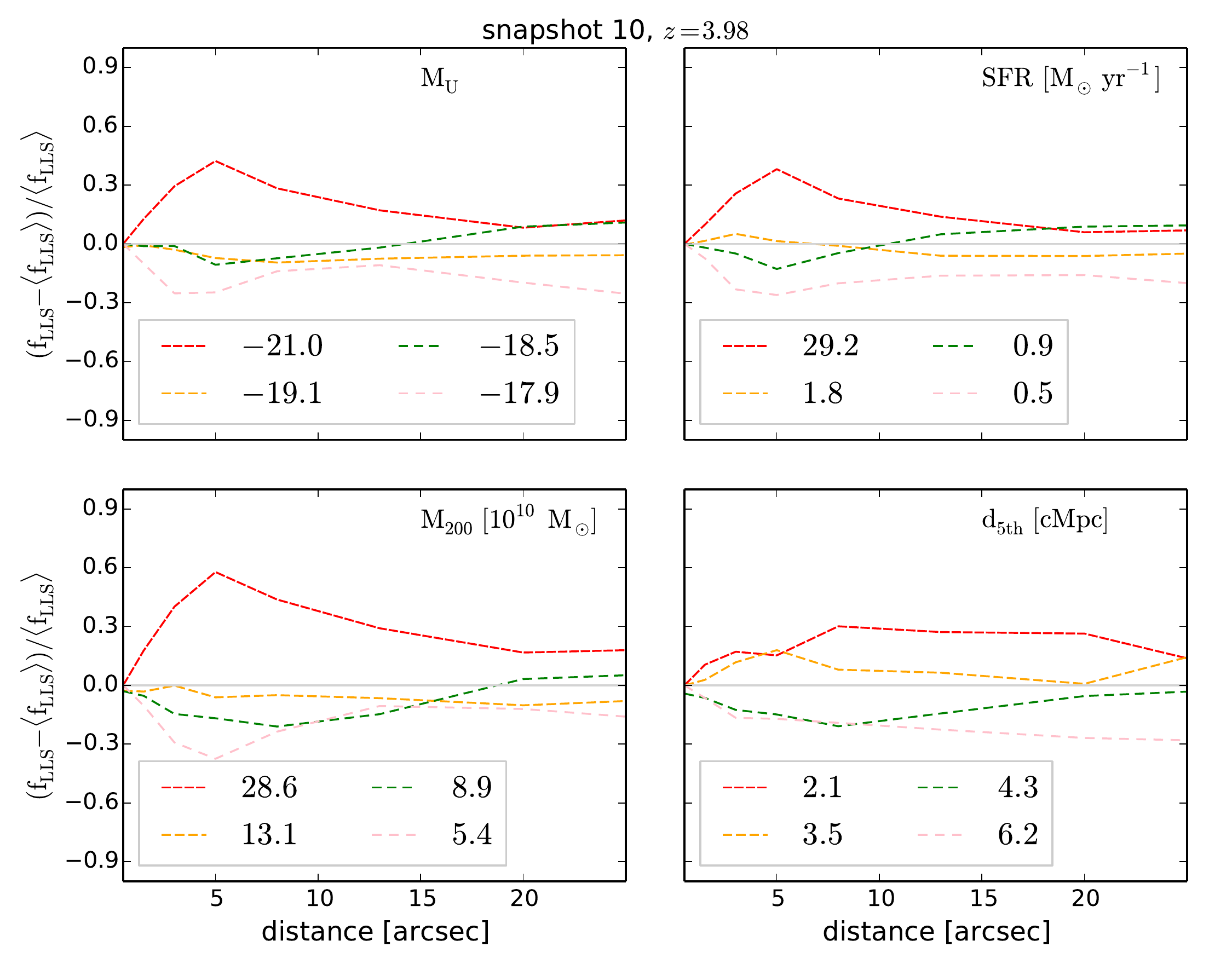}
\includegraphics[width=.49\textwidth]
{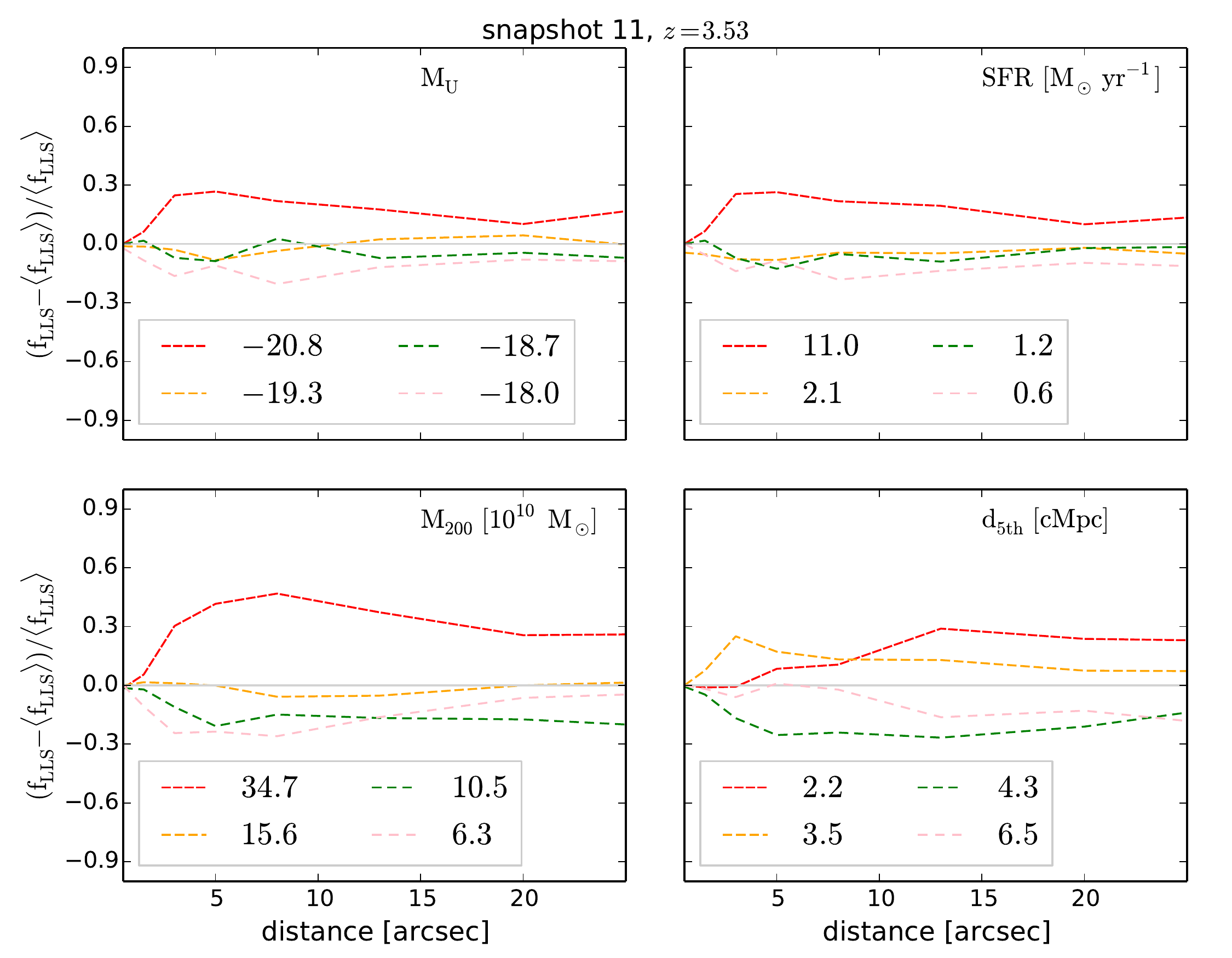}
\includegraphics[width=.49\textwidth]
{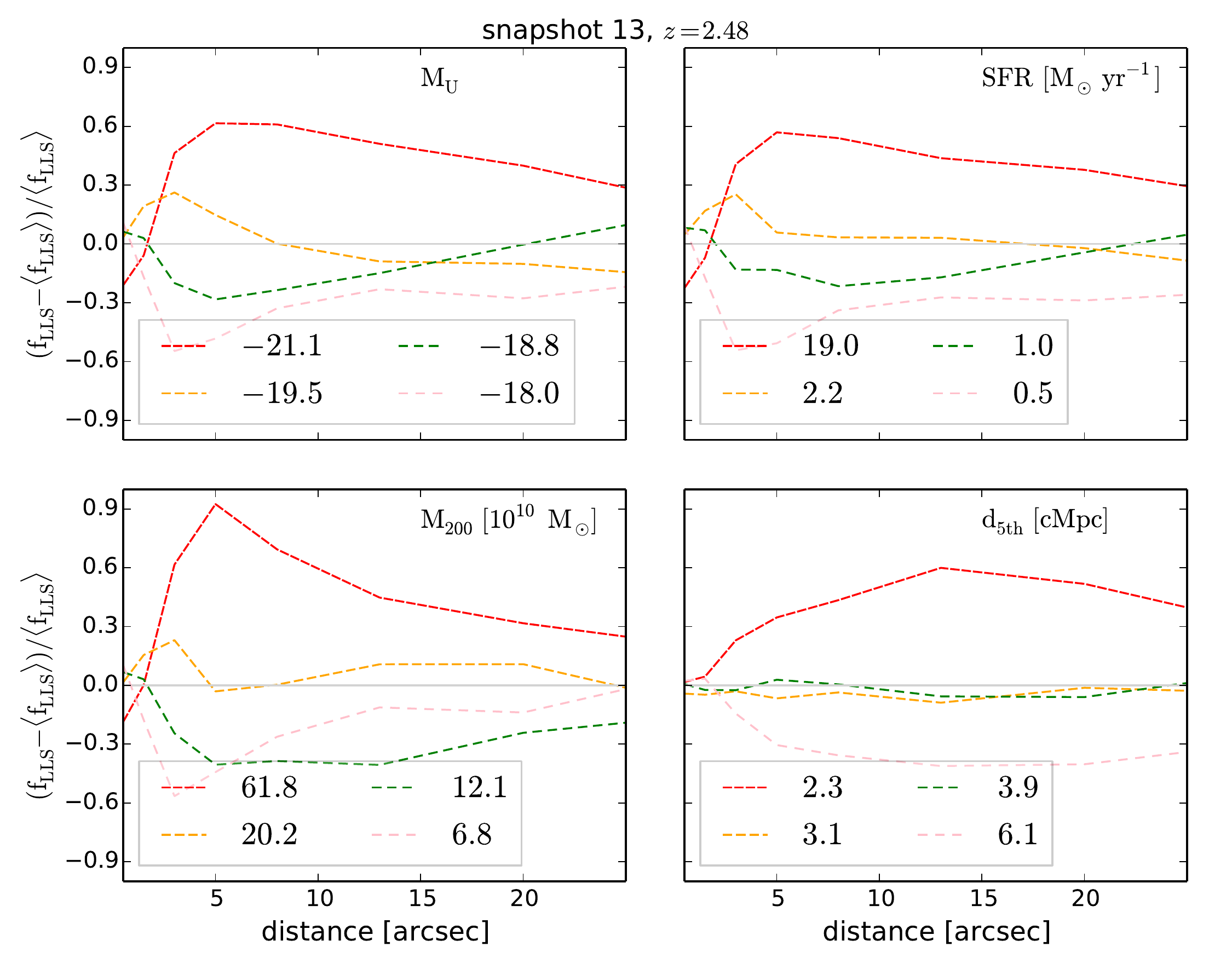}
\includegraphics[width=.49\textwidth]
{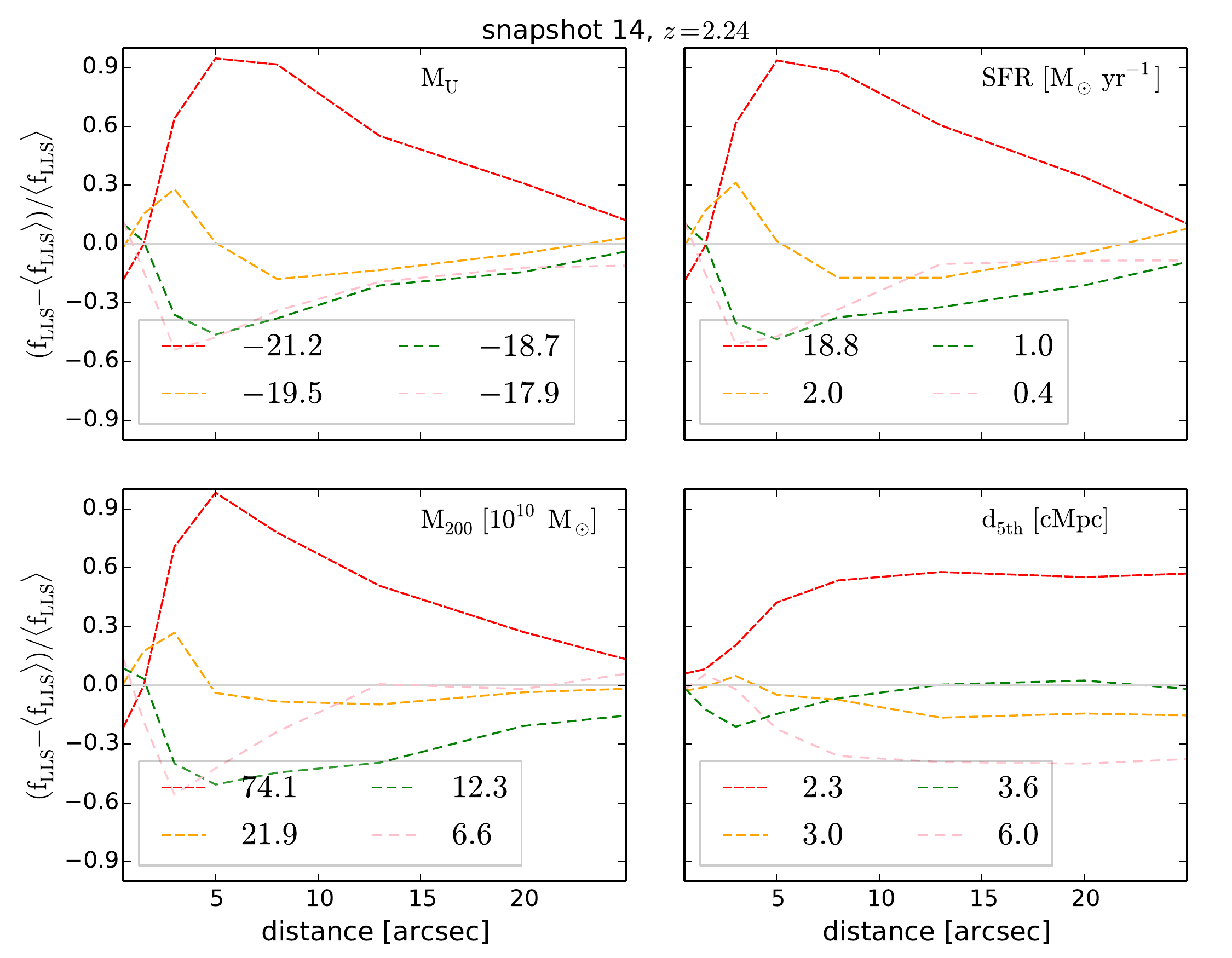}
   \caption{Same as Figure \ref{lls12} for the rest of the EAGLE mock cubes generated.} \label{LLSsnaps}
\end{figure*}

We explore how relevant are different galaxy properties on the expected \flls\ around galaxies in our simulations at different redshifts, which can be seen in Fig. \ref{LLSsnaps} (see Fig. \ref{lls12} for $z=3.02$). For $z=5$ we observe a clear positive correlation with $\mathrm{M_{200}}$ at small scales and with $\mathrm{d_{5th}}$ at large scales. For $z<4.5$ all properties are correlated with \flls\ with increasing significance up to $z=2.2$. At $z=2.2$, there is a sharp decrease in the correlation for all galaxy properties except $\mathrm{d_{5th}}$. As discussed in Sec. \ref{sec:galprops}, $\mathrm{d_{5th}}$ is an easy property to estimate, potentially useful for IGM studies targeting regions with homogeneous galaxy surveys available at those redshifts.

\label{lastpage}

\end{document}